\RequirePackage{ifpdf}
\documentclass[11pt,letterpaper]{JHEP3}
\usepackage{epsfig}
\usepackage[latin1]{inputenc}
\usepackage{bbold,amsfonts}
\usepackage{graphicx}
\usepackage{amssymb,amsmath}
\usepackage{fancybox,framed}
\usepackage{dsfont}
\usepackage{mathtools}
\usepackage{braket}
\usepackage{slashed}

\author{Lorenzo~Bianchi$^{a}$, Marco S. Bianchi$^{b}$\\\\
$^{a}$ Institut f\"ur Physik,
Humboldt-Universit\"at
zu Berlin\\ Newtonstra{\ss}e 15, 12489 Berlin, Germany
\\
$^{b}$
Centre for Research in String Theory,
School of Physics and Astronomy\\
Queen Mary University of London,
Mile End Road, London E1 4NS, UK \\
\qquad\\
E-mail: \email{lorenzo.bianchi@physik.hu-berlin.de, m.s.bianchi@qmul.ac.uk}
}

\abstract{We compute the one-loop correction to the dispersion relations of the excitations of the $AdS_4\times \mathbb{CP}^3$ sigma model dual to ABJM theory, expanded around the cusp background.
The results parallel those of ${\cal N}=4$ SYM. As in that case, the dispersion relations are compatible with the predictions from the Bethe ansatz for the GKP string, though showing some known discrepancies on which we comment.
}

\preprint{May 2015\\HU-EP-15/22\\QMUL-PH-15-09}

\title{Quantum dispersion relations for the $AdS_4\times CP^3$ GKP string.}


\csname @addtoreset\endcsname{equation}{section}

\def\bseq{\begin{subequation}}  % = 1a 1b
\def\eseq{\end{subequation}}
\def\bsea{\begin{subeqnarray}}  % = 1.1a 1.1b
\def\esea{\end{subeqnarray}}
\def\backslashedpa#1{#1\!\!\!\backslash}

\newcommand{\beq}{\begin{equation}}
\newcommand{\bea}{\begin{eqnarray}}
\newcommand{\eea}{\end{eqnarray}}
\newcommand{\eeq}{\end{equation}}

\renewcommand{\d}{\delta}
\newcommand{\pa}{\partial}
\newcommand{\g}{\gamma}

\newcommand{\e}{\epsilon}

\renewcommand{\l}{\lambda}

\newcommand{\vf}{\varphi}

\newcommand{\p}{\pi}

\renewcommand{\O}{\Omega}

\newcommand{\ppp}{\text{p}}

\def\beq{\begin{equation}}
\def\eeq{\end{equation}}
\def\bea{\begin{eqnarray}}
\def\eea{\end{eqnarray}}
\def\Tr{\mathrm{Tr}}

\def\g{\gamma}

\def\d{\delta}
\def\e{\epsilon}

\def\l{\lambda}

\def\vf{\varphi}

\declareslashed{}{\backslash}{0}{0}{\Omega}
\declareslashed{}{\backslash}{0}{0}{O}
\begin{document}

\allowdisplaybreaks

\section{Introduction}

The AdS/CFT correspondence \cite{Maldacena:1997re} identifies the spectrum of anomalous dimensions of local operators in a conformal field theory with the energy spectrum of strings in an $AdS$ background.
In its prototypical instance, the spectrum of the superstring in $AdS_5\times S^5$ matches that of local operators in the planar limit of ${\cal N}=4$ SYM theory.
Moreover, the spectral problem of the latter is conjectured to be integrable \cite{Minahan:2002ve,Beisert:2003yb}. This allows to interpolate exactly the anomalous dimensions data from weak to strong coupling, which in turn enables to compare them with the strong coupling predictions from string theory.

A well studied example of this state-operator correspondence and the underlying integrability of the spectral problem is the sector of twist-two operators in planar ${\cal N}=4$ SYM.
These operators have the schematic form $\Tr (Z D_{+}^S Z)$, where $Z$ are complex scalars of ${\cal N}=4$ SYM theory, $D_{+}$ is the covariant light-cone derivative and $S$ is the spin of the operator. Such operators are closed under renormalization, i.e. they only mix among themselves, thus providing a closed sector.
At weak coupling, in the large spin limit, the anomalous dimension of twist-two operators scales logarithmically~\cite{Belitsky:2006en}, with a slope given by the cusp anomalous dimension $\Gamma_{cusp}$, which is a non-trivial function of the 't Hooft coupling. 
In the strong regime, such twist-two operators are conjectured to be dual to a folded string spinning around its center of mass in $AdS_3 \subset AdS_5$ \cite{Gubser:2002tv}. Its rather complicated solution was spelled out in \cite{Frolov:2002av}.
In the large spin limit this solution simplifies and corresponds to a homogeneous long string stretching up to the boundary of $AdS_5$ \cite{Gubser:2002tv} (that we shall also refer to as the GKP string).
The cusp anomalous dimension $\Gamma_{cusp}$ represents its uniformly distributed energy\footnote{In general subleading corrections at strong coupling in the large spin limit were considered in \cite{Beccaria:2010ry}.}.
This ubiquitous quantity emerges also in the gauge theory as the coefficient of the UV divergence of a cusped, light-like Wilson loop~\cite{Polyakov:1980ca,Korchemsky:1985xj,Korchemsky:1987wg}.
From the string perspective, the worldsheet area corresponding to the spinning string solution is equivalent to that of a null cusp \cite{Kruczenski:2002fb,Alday:2007mf}, thus explaining the appearance of the same function in two naively unrelated contexts.

The interpolation between the weak and strong coupling descriptions can be made precise thanks to integrability, by means of a set of Bethe equations \cite{Beisert:2005fw}.
In particular, the integrability based BES equation \cite{Beisert:2006ez} allows in principle to compute $\Gamma_{cusp}$ to any desired order in both regimes. 
At strong coupling this expansion corresponds to considering the classical energy of the string in the null cusp background and adding its quantum corrections that can be computed by summing over all excitations about the vacuum.
Focussing on the individual excitations, they are interpreted at weak coupling as describing the spectrum of anomalous dimensions of higher twist operators, obtained from twist-two ones by insertion of extra fields.
Integrability allows to extract exact results for the excitations as well.
In particular, from the Bethe ansatz description, not only the energy but also the momentum of the excitations can be computed, so that their exact dispersion relations were derived in \cite{Basso:2010in} (previous results appeared in \cite{Belitsky:2006en,Eden:2006rx,Freyhult:2007pz,Bombardelli:2008ah}).
Moreover, their scattering amplitudes have also been studied in \cite{Basso:2013pxa,Fioravanti:2013eia,Fioravanti:2015dma}.

Recently, additional interest on the GKP string excitations was triggered by the OPE approach to light-like polygonal Wilson loops and scattering amplitudes \cite{Alday:2010ku,Gaiotto:2011dt,Basso:2013vsa,Basso:2013aha,Basso:2014koa,Basso:2014jfa,
Basso:2014nra,Basso:2014hfa}, where the GKP scattering factors appear as crucial building blocks.
\bigskip

The $AdS_4/CFT_3$ correspondence offers another setting where to study analogous problems.
Indeed, on the one hand there exists a well defined correspondence between a three-dimensional conformal field theory (the ABJM model \cite{Aharony:2008ug}) and a dual string theory, namely type IIA on $AdS_4\times \mathbb{CP}^3$.
On the other hand the ABJM theory was conjectured to be integrable \cite{Minahan:2008hf,Gaiotto:2008cg,Grignani:2008is,Bak:2008cp,Gromov:2008qe} and its Bethe equations to be related to those of ${\cal N}=4$ SYM in a precise and simple way.
Interestingly, the all loop Bethe ansatz for the ABJM theory features an interpolating function of the coupling $h(\lambda)$, which is trivial for ${\cal N}=4$ SYM.
Therefore, ABJM theory offers a rather similar, though somewhat different environment, where to put the AdS/CFT and integrability machinery in action.

The spinning string problem has been extensively studied in the ABJM theory as well \cite{Alday:2008ut,McLoughlin:2008ms,Krishnan:2008zs,McLoughlin:2008he,Forini:2012bb}, where the one-loop corrections to its energy were computed.
These correspond to the cusp anomalous dimension of ABJM at strong coupling, which can be computed from integrability via a BES equation, which is the same as for ${\cal N}=4$ SYM, up to the presence of the effective coupling $h(\lambda)$. Therefore the comparison of the two results yields the strong coupling expansion of the interpolating function.

As in ${\cal N}=4$ SYM attention was devoted not only to the sum over all the excitations on top of the spinning string ground state, but also to the excitations themselves.
In particular, the model governing the low-energy excitations of the $AdS_4\times \mathbb{CP}^3$ GKP vacuum in the Alday-Maldacena limit \cite{Alday:2007mf} was pointed out in \cite{Bykov:2010tv}.
It belongs to a larger group of models which were considered in \cite{Basso:2012bw}, where it was argued that they are integrable.

In \cite{Basso:2013pxa} the asymptotic Bethe ansatz for the ABJM GKP string was studied, from which the authors argued the dispersion relations and S-matrix for the excitations. One difference with respect to $\mathcal{N}=4$ SYM is that there is no closed subsector with derivatives and scalar fields only and the simplest set of operators dual to the spinning string solution is of the form $\Tr(D_+...D_+Y^1D_+...D_+ \psi_{4+}^\dagger D_+\psi^1_+D_+Y^\dagger_4)$ built out of bifundamental matter fields $(Y^1,\psi_+^1), (Y_4^\dagger,\psi_{4+}^\dagger)$. To identify the GKP vacuum one has to look for the state with the lowest possible twist. In this case it is provided by a twist-one\footnote{Notice that both scalar fields and fermions in three dimensions have twist 1/2.} operator (to be compared to the twist-two vacuum of $\mathcal{N}=4$) containing two bifundamental matter fields and a large number $S$ of covariant derivatives. In this picture the lowest lying excitations are the twist-1/2 matter fields which 
transform in the $\mathbf{4}$ and $\mathbf{\bar{4}}$ representations of $SU(4)$. They are accompanied by twist-1/2 fermions in the $\mathbf{6}$ representation of $SO(6)$ and a tower of twist-$\ell$ 
excitations, neutral under $SU(4)$, corresponding to the transverse component of the gauge field \cite{Basso:2013pxa}. Despite the qualitative difference with respect to $\mathcal{N}=4$ SYM, the similarity of the two integrable models predicts closely related dispersion relations for the excitations in the two theories.
\bigskip

In the present paper we provide an explicit computation of the one-loop corrections to the dispersion relations of the excitations of the ABJM GKP string at strong coupling, via sigma model perturbation theory.
This constitutes a strong verification of the predictions of integrability at the quantum level.
In order to make the computation feasible, we use the Lagrangian of the $AdS_4\times \mathbb{CP}^3$ superstring in the light-cone gauge, expanded about the light-like cusp background.
Indeed on the one hand this setting was proved to be equivalent to that of the GKP spinning string, on the other hand the light-cone gauge Lagrangian turns out to be more tractable and suitable for perturbative computations.

In the context of $AdS_5\times S^5$ this approach, initiated in \cite{Metsaev:2000yu,Metsaev:2000yf}, allowed for the computation of the two-loop free energy \cite{Giombi:2009gd}, which is interpreted as the cusp anomalous dimension via the AdS/CFT correspondence. In a following paper \cite{Giombi:2010bj} the quantum dispersion relations for the excitations of this model were computed at one loop and compared to the integrability predictions of \cite{Basso:2010in}. A refined analysis on the stability of the heaviest scalar mode, the binding energy of gauge excitations bound states and the corrections to the dispersion relation of massless states was presented in \cite{Zarembo:2011ag}.

A parallel program was started for the $AdS_4\times \mathbb{CP}^3$ case. In \cite{Uvarov:2009hf,Uvarov:2009nk,Uvarov:2011zz} the light-cone gauge superstring Lagrangian was derived. In \cite{Bianchi:2014ada} it was expanded around the null cusp background and the two-loop correction to the cusp anomalous dimension of ABJM at strong coupling was determined. This also provided a non-trivial check on the conjectured exact expression of the interpolating function of the 't Hooft coupling $h(\lambda)$ \cite{Gromov:2014eha}, featured in integrability based computations in ABJM.
As for the $AdS_5\times S^5$ case, the light-cone gauge Lagrangian offers an efficient setting for computing the dispersion relations of the excitations on the cusp background.
In this paper we determine their quantum corrections at one loop and compare them with the predictions from the Bethe ansatz of \cite{Basso:2013pxa}.
Here we summarize our results. The string theory spectrum at $\l\to \infty$ consists of
\begin{align}
 AdS_3\textup{ transverse mode }(\varphi):	\qquad&m_\varphi^2=4\\
 AdS_4 \textup{ outside } AdS_3\ (x):		\qquad&m_x^2=2\\
 \mathbb{CP}^3\ (\{z^a,\bar z_a\}, a=1,2,3):	\qquad&m_z^2=0\\
 \textup{Massive fermions } (\eta^a,\theta^a):	\qquad&m_{\eta^a}^2=m_{\theta^a}^2=1\\
 \textup{Massless fermions } (\eta^4,\theta^4):	\qquad&m_{\eta^4}^2=m_{\theta^4}^2=0
\end{align}
We find the following quantum corrections to the dispersion relations and masses of those excitations, which can be compared to the results of \cite{Giombi:2010bj} by replacing $h(\l)\to \frac{\sqrt{\l}}{4\,\pi}$
\begin{align}\label{eq:result}
 &E^2(\ppp,\l)=\left[\ppp^2+m^2+\frac{q}{h(\l)}+\mathcal{O}(\l^{-1})\right]\left[1+\frac{c\, \ppp^2+d }{h(\l)}+\mathcal{O}(\l^{-1})\right]\\
 &\begin{array}{lllll}
 q_\varphi=0 		\quad&q_x=-\frac14  	\quad&q_z=0  			\quad&q_{\eta^a}=q_{\theta^a}=0  		\quad&q_{\eta^4}=q_{\theta^4}=0\\
 c_\varphi=-\frac18 	\quad&c_x=-\frac14 	\quad&c_z=-\frac{11}{12\,\p}	\quad&c_{\eta^a}=c_{\theta^a}=-\frac12  	\quad&c_{\eta^4}=c_{\theta^4}=-\frac{7}{4\,\p}\\
 d_\varphi=0 		\quad&d_x=0  		\quad&d_z=0  			\quad&d_{\eta^a}=d_{\theta^a}=0  		\quad&d_{\eta^4}=d_{\theta^4}=\frac{1}{\p}
\end{array}
\end{align}

For massive modes the quantum dispersion relations of the string description are in agreement with those of integrability. 
On the other hand, an identification of the massless modes and their dispersion relations turns out to be rather problematic in a perturbative approach.
A similar mismatch has been already highlighted in \cite{Zarembo:2011ag} in the context of $AdS_5\times S^5$. In that case the five massless scalars of the superstring sigma model are not in direct correspondence with the six massless degrees of freedom of the integrability description (and the low energy $O(6)$ sigma model).
Moreover it was pointed out that the perturbative expansion of their dispersion relations as computed from the superstring sigma model is ill-defined because of IR divergences.
We encounter a parallel mismatch for massless modes in the $AdS_4\times \mathbb{CP}^3$ case, where the four complex scalars coupled to a Dirac fermions of the perturbative string sigma model are not in correspondence with the excitations of the integrability description. 
The reason why this is so can be traced to the nonperturbative dynamics of the low-energy excitations, which describe the Bykov model.
This is a $\mathbb{CP}^3$ sigma model coupled to a Dirac fermion. Nonperturbatively, the fermion condensates, spontaneously breaking the $U(1)$ symmetry of the model and giving a mass to its gauge field.
This screens long range interactions and prevents scalars from confining (which happens in bosonic $\mathbb{CP}^N$ models).
The massless fermion confines and hence does not constitute an asymptotic degree of freedom of the theory.
Eventually the spectrum consists of 4+4 excitations (spinons) acquiring a nonperturbative mass, which are identified with the holes of the integrability description.

Furthermore, as was pointed out in \cite{Zarembo:2011ag} for the $AdS_5\times S^5$ case, the heaviest scalar mode of the sigma model is not present in the integrability analysis as an elementary excitation. A similar issue is encountered in the $AdS_4\times \mathbb{CP}^3$ setting and in the following we comment on the interpretation of such a mode. 

Finally, for gauge excitations integrability predicts the formation of bound states and allows to compute their binding energy exactly. Following a similar analysis to \cite{Zarembo:2011ag}, we estimate this binding energy from the non-relativistic limit of the scattering amplitude between these modes, which is compatible with the integrability result in the static approximation.

The paper is organized as follows: in Section \ref{sec:Lagrangian} we review the light-cone gauge Lagrangian of $AdS_4\times \mathbb{CP}^3$ and its expansion up to fourth order in the fields, about the null cusp background.
Section \ref{sec:dispersion} contains the computation of the one-loop correction to the dispersion relations of the excitations. 
In section \ref{sec:comparison} we comment the results and compare them to those of ${\cal N}=4$ SYM and integrability.
Section \ref{heavy} is devoted to a discussion on the r\^ole of the heaviest scalar.
The possibility for the excitations to form bound states is explored in \ref{sec:boundstates} in the non-relativistic limit.
Finally in the conclusions \ref{sec:conclusions} we summarize our findings.
In appendix \ref{app:lagr_exp} we spell out the details of the expanded light-cone Lagrangian up to fourth order in the excitations.
In appendix \ref{app:lagWZ} we provide an alternative form of the Lagrangian, namely the WZ type which we use in appendix \ref{sec:lagdirac} to rewrite it using Dirac fermions.

\section{The light-cone gauge Lagrangian}\label{sec:Lagrangian}

We review the light-cone gauge fixed action of $AdS_4\times \mathbb{CP}^3$ and its expansion about the null cusp background.

\subsection{The light-cone gauge $AdS_4\times \mathbb{CP}^3$ Lagrangian}
The starting point is the $AdS_4\times \mathbb{CP}^3$ metric, which we take to be
\begin{equation}\label{eq:metric}
ds^2_{10} = R^2 \left( \frac14\, ds^2_{AdS_4} + ds^2_{\mathbb{CP}^3} \right)\,,
\end{equation}
where $R$ is the $\mathbb{CP}^3$ radius.
The $AdS_4$ metric in the Poincar\'e patch reads
\begin{eqnarray}
ds^2_{AdS_4} &=& \frac{dw^2 + dx^+ dx^- + dx^1dx^1}{w^2} \qquad\qquad x^{\pm} \equiv x^2 \pm 
x^0\,,%\\
%ds^2_{\mathbb{CP}^3}&=&g_{MN}\,dz^M dz^N\,\qquad\qquad M=1,...,6~.
\end{eqnarray}
where $x^\pm$ are the light-cone coordinates, $x^m=(x^0,x^1,x^2)$ span the boundary of $AdS_4$ and $w\equiv e^{2\varphi}$ is the radial direction.
We parametrize $\mathbb{CP}^3$ with complex variables $z^a$ and $\bar z_a$, 
transforming in the $\mathbf{3}$ and $\mathbf{\bar 3}$ of $SU(3)$ respectively,
\begin{equation}
ds^2_{\mathbb{CP}^3} = g_{MN}\, z^M\, z^N = g_{ab}\, dz^a\, dz^b + g^{ab}\, d\bar z_a\, d\bar z_b + 2\, 
g_a^{\phantom{a}b}\, dz^a\,d\bar z_b\,,
\end{equation}
where 
\begin{align}
& g_{ab} = \frac{1}{4|z|^4} \left( |z|^2 - \sin^2{|z|} + \sin^4{|z|} \right)\bar z_a\,\bar z_b\, , \qquad g^{ab} = \frac{1}{4|z|^4} \left( |z|^2 - \sin^2{|z|} + \sin^4{|z|} \right) z^a\, z^b\, , & \nonumber\\
& g_a^{\phantom{a}b} = \frac{\sin^2{|z|}}{2|z|^2}\, \delta_a^b + \frac{1}{4|z|^4} \left( |z|^2 - \sin^2{|z|} - \sin^4{|z|} \right)\bar z_a\, z^b \qquad \text{and}\quad |z|^2 \equiv z^a\, \bar z_a\,. &
\end{align}
In addition to the bosonic fields which parametrize the $AdS_4\times \mathbb{CP}^3$ metric, the Lagrangian contains the fermionic coordinates $\eta_a$ and $\theta_a$ with index $a=1,2,3$, transforming in the fundamental representation of $SU(3)$.
They stem for the 24 unbroken supersymmetries of the background, out of the original 32 of type IIA supergravity.
The eight supersymetries broken by the background manifest themselves with the fermions $\eta_4$, $\theta_4$. Complex conjugation of fermions is achieved by raising the indices $\{a,4\}$. Fermions with upper index $a$ ($\eta^a,\theta^a$) transform in the anti-fundamental of $SU(3)$\footnote{Compared to \cite{Uvarov:2009hf, Uvarov:2009nk, Bianchi:2014ada}, here we omit the bars to indicate the complex conjugation of fermions. This does not generate any confusion and simplifies several expressions.}.

The $\kappa$-symmetry light-cone gauge-fixed Lagrangian of~\cite{Uvarov:2009hf, Uvarov:2009nk} takes the form
\begin{align}\label{eq:lagrangian}
S &= -\frac{T}{2} \int\, d\tau\, d\sigma\, L \qquad\qquad \\ %T = \frac{R^2}{2\pi\alpha'}\\ 
L &= \gamma^{ij}\Big[\frac{e^{-4\varphi}}{4} \left(\partial_ix^+\partial_jx^-+
\partial_ix^1\partial_jx^1 \right) + \partial_i\varphi
\partial_j\varphi + g_{MN}\partial_i z^M \partial_j z^N\nonumber\\&\phantom{=\gamma^{ij}}+
e^{-4\varphi}\left(\partial_ix^+\varpi_j+\partial_ix^+\partial_jz^M h_M+e^{-4\vf} B\, \partial_ix^+\partial_jx^+\right)\Big]\nonumber\\
&-2\, \varepsilon^{ij}e^{-4\varphi}\left({\omega}_i\partial_jx^++e^{-2\varphi}C\, \partial_ix^1\partial_jx^+
+\partial_ix^+\partial_jz^M\ell_M\right)\nonumber \quad ,
\end{align}
in terms of
\begin{align}\label{eq:pieces}
\varpi_i&=i\left(\partial_i\theta_a{\theta}^a-\theta_a\partial_i{\theta}^a
+\partial_i\theta_4{\theta}^4-\theta_4\partial_i{\theta}^4
+\partial_i\eta_a{\eta}^a-\eta_a\partial_i{\eta}^a
+\partial_i\eta_4{\eta}^4-\eta_4\partial_i{\eta}^4\right) \, ,\\ 
\omega_i&=\hat{\eta}_a\hat{\partial_i}{\theta}^a
+\hat{\partial_i}\theta_a\hat{{\eta}}^a+\frac12\left(\partial_i\theta_4{\eta}^4-\partial_i\eta_4
{\theta}^4+\eta_4\partial_i{\theta}^4-\theta_4
\partial_i{\eta}^4\right) \, , \\ \label{eq:B}
B&=8\,  \left[(\hat{\eta}_a\hat{{\eta}}^a)^2+\varepsilon_{abc}\hat{{\eta}}^a\hat{{\eta}}^b
\hat{{\eta}}^c{\eta}^4+\varepsilon^{abc}
\hat{\eta}_a\hat{\eta}_b\hat{\eta}_c\eta_4+2
\eta_4{\eta}^4 \left(\hat{\eta}_a\hat{{\eta}}^a-\theta_4{\theta}^4\right)\right] \, , \\ 
C&=2\, \hat{\eta}_a\hat{{\eta}}^a+
\theta_4{\theta}^4+\eta_4{\eta}^4 \, ,\\ 
h_M&=2\, \left[\Omega^a_M\varepsilon_{abc}\hat{{\eta}}^b\hat{
{\eta}}^c-\Omega_{aM}\varepsilon^{abc}\hat{\eta}_b
\hat{\eta}_c + 2\left(\Omega_{aM}\hat{{\eta}}^a
{\eta}^4-\Omega^a_M\hat{\eta}_a\eta_4\right) + 2
\left(\theta_4{\theta}^4+\eta_4{\eta}^4\right)
\tilde{\Omega}_{a\ M}^{\ a}\right]\, , \\ \label{eq:ellM}
\ell_M&=2\, i\, \left[\Omega_{aM}\hat
{{\eta}}^a{\theta}^4+\Omega^a_M\hat{\eta}_a\theta_4
+ \left(\theta_4{\eta}^4-\eta_4{\theta}^4\right)
\tilde{\Omega}_{a\ M}^{\ a}\right] 
\end{align}
A few explanations of the objects appearing in the Lagrangian are in order.
The string tension $T$, including the anomalous radius shift, is given by \cite{Bergman:2009zh}
\begin{equation}\label{eq:tension}
T = \frac{R^2}{2\pi} = 2\sqrt{2\left(\lambda-\frac{1}{24}\right)}\,,
\end{equation}
where $\lambda\equiv \frac{N}{k}$ is the 't Hooft coupling.
The parameters $N$ and $k$ are the rank and the level of the Chern-Simons $U(N)_k\times U(N)_{-k}$ ABJM field theoretical description at weak coupling and correspond to the units of four- and two-form flux respectively in the string theory dual in the strong regime.
Corrections from the anomalous shift begin to affect terms at $\lambda^{-1}$ order in perturbation theory and can be disregarded in this paper where we work at first order, hence we approximate $T=2\sqrt{2\lambda}$ throughout this paper.
When presenting results we also use the interpolating function $h(\lambda)$ in order to make a more direct contact with predictions from integrability.
To one loop order for $\lambda \gg 1$
\begin{equation}\label{eq:h}
h(\lambda) = \sqrt{\frac{\lambda}{2}} = \frac{T}{4}\,.
\end{equation}

The $\Omega^a_M$ and $\Omega_{aM}$ appearing in the Lagrangian are the complex vielbein 
of $\mathbb{CP}^3$, $ds^2_{\mathbb{CP}^3} = \Omega^a_M \Omega_{aN}\, dz^M\, 
dz^N$, and $\tilde\Omega_a^{\phantom{a}a} $ is associated to a one-form corresponding to the fiber direction of $S^7$, when dimensionally reducing from the supermembrane action in the $D=11$ $AdS_4\times \mathbb{CP}^3$ background. 
Explicitly,
\begin{align}\label{eq:forms}
%\Omega_{a}^{\phantom{a}b}&=i\frac{(1-\cos{|z|})}{|z|^2}(\bar z_adz^b-d\bar z_az^b)-i\bar z_az^b\frac{(1-\cos{|z|})^2}{2|z|^4}(dz^c\bar z_c-z^cd\bar z_c),\\
\Omega_{a}&=d\bar z_a\frac{\sin{|z|}}{|z|}+\bar z_a\frac{\sin{|z|}(1-\cos{|z|})}{2|z|^3}(dz^c\bar z_c-z^cd\bar z_c)+\bar z_a\left(\frac{1}{|z|}-\frac{\sin{|z|}}{|z|^2}\right)d|z|,\\
\Omega^{a}&=dz^a\frac{\sin{|z|}}{|z|}+z^a\frac{\sin{|z|}(1-\cos{|z|})}{2|z|^3}(z^cd\bar
z_c-dz^c\bar
z_c)+z^a\left(\frac{1}{|z|}-\frac{\sin{|z|}}{|z|^2}\right)d|z|.
\end{align}
and 
\begin{equation}\label{omtildecc}
\tilde \Omega_a^{\phantom{a}a} = i\, \frac{\sin^2|z|}{|z|^2} \left( dz^a\, \bar z_a - z^a\, d\bar z_a 
\right)\,.
\end{equation}
Finally, the hatted fermions appearing in the Lagrangian are just a rotation 
\begin{equation}\label{eq:T}
\hat\eta_a = T_a^{\phantom{a}b}\, \eta_b + T_{ab}\,\eta^b\, , \qquad\qquad \hat{\eta}^a 
= T^a_{\phantom{a}b}\, \eta^b + T^{ab}\,\eta_b\,,
\end{equation}
by matrices $T$
\begin{equation}\label{matrixT}
{T_{\hat{a}}}^{\hat b}=\left(
 \begin{array}{cc}
  T_a^{\phantom{a}b} & T_{ab} \\
  T^{ab} & T^a_{\phantom{a}b}
 \end{array}\right) %= \left(
% \begin{array}{cc}
%  0 & i\,\varepsilon_{acb}\, z^c \\
 % -i\,\varepsilon^{acb}\, \bar z_c & 0
 %\end{array}\right)= 
= \left(
 \begin{array}{cc}
  \delta_a^b\, \cos{|z|} + \bar z_a\, z^b\, \frac{1-\cos{|z|}}{|z|^2} & i\,\varepsilon_{acb}\, z^c\, \frac{\sin{|z|}}{|z|} \\
  -i\,\varepsilon^{acb}\, \bar z_c\, \frac{\sin{|z|}}{|z|} & \delta^a_b\, \cos{|z|} + z^a\, \bar z_b\, \frac{1-\cos{|z|}}{|z|^2}
 \end{array}\right)\,.
\end{equation} 
The bosonic world-sheet local symmetry is fixed with a ``modified'' conformal gauge
 \begin{equation}\label{eq:nonconf}
\gamma^{ij} = {\rm diag}\left(-e^{4\varphi}, e^{-4\varphi}\right)\,,
\end{equation}
and imposing the light-cone gauge condition 
\begin{equation}\label{eq:lcdir}
x^+=p^+\,\tau\,,\qquad p^+={\rm const}\,.
\end{equation}

\subsection{The null cusp vacuum and fluctuations}\label{sec:expansion}
The Lagrangian \eqref{eq:lagrangian} admits a vacuum corresponding to a null cusp at the boundary of $AdS_4$ \cite{Kruczenski:2002fb, Giombi:2009gd}
\begin{eqnarray}\label{eq:background}
& w \equiv e^{2\varphi} = \displaystyle\sqrt{\frac{\tau}{\sigma}} \qquad\qquad
x^{1} = 0 &\nonumber \\
& x^{+} =\tau \qquad\qquad
x^{-} =-\displaystyle\frac{1}{2\sigma}  \qquad\qquad
z^{a} = \bar z_a = 0~. &
\end{eqnarray} 
We can expand around this background setting
\begin{eqnarray}\label{eq:fluctuations}
& x^1 = 2\, \displaystyle\sqrt{\frac{\tau}{\sigma}} \tilde x \qquad\qquad w = \displaystyle\sqrt{\frac{\tau}{\sigma}}\, \tilde w \qquad\qquad \tilde w = e^{2\tilde\varphi}  & \nonumber\\
& z^a = \tilde z^a \qquad\qquad \bar z_a = \tilde{\bar z}_a \qquad\qquad a=1,2,3 & \nonumber\\
& \eta = \displaystyle\frac{1}{\sqrt{2\, \sigma}}\, \tilde \eta \qquad\qquad \theta = \displaystyle\frac{1}{\sqrt{2\,\sigma}}\, \tilde \theta\,. & 
\end{eqnarray}
Next we rotate the Lagrangian \eqref{eq:lagrangian} to Euclidean signature and redefine the worldsheet coordinates as $t =2\, \log \tau$ and $s=2\, \log\sigma$, so that the fluctuation Lagrangian reads
\begin{equation}\label{eq:action}
S_E = \frac{T}{2}\, \int dt\, ds\, {\cal L}\qquad,\qquad
{\cal L} = {\cal L}_B + {\cal L}_F^{(2)} + {\cal L}_F^{(4)}\,,
\end{equation}
where
\begin{align}
 {\cal L}_B& = \left( \partial_t \tilde x +  \tilde x \right)^2 + \frac{1}{\tilde w^4} \left( \partial_s \tilde x - \tilde x \right)^2
+ \tilde w^2\, \left(\partial_t \tilde \varphi \right)^2 + \frac{1}{\tilde w^2}\, \left(\partial_s \tilde \varphi \right)^2 + \frac{1}{4} \left( \tilde w^2 + \frac{1}{\tilde w^2} \right) 
\nonumber\\& 
+ \tilde w^2 \, \tilde g_{MN}\, \partial_t \tilde z^M\, \partial_t \tilde z^N + \frac{1}{\tilde w^{2}}\, \tilde g_{MN}\, \partial_s \tilde z^M\, \partial_s \tilde z^N \label{boslag}
\\
{\cal L}_F^{(2)} &= i\Big[  \partial_t {\tilde\theta}_a {\tilde\theta}^a - {\tilde\theta}_a\partial_t{\tilde\theta}^a
+ \partial_t{\tilde\theta_4}{\tilde\theta}^4 - {\tilde\theta_4}\partial_t
{\tilde\theta}^4 + \partial_t{\tilde\eta_a}{\tilde\eta}^a - {\tilde\eta_a}\partial_t{\tilde\eta}^a 
+ \partial_t\tilde{\eta_4}\tilde{\eta}^4 - \tilde{\eta_4}\partial_t
\tilde{\eta}^4 \Big] \nonumber\\&
+\frac{2i}{\tilde w^2}\Big[\hat{\tilde\eta}_a \left(\hat{\partial_s}{\tilde\theta}^a - \hat{ \tilde\theta}^a \right)
+ \left(\hat{\partial_s}\tilde\theta_a - {\hat {\tilde\theta}}_a \right) \hat{\tilde{\eta}}^a + \frac12 \left(\partial_s\tilde\theta_4{\tilde\eta}^4 - \partial_s\tilde\eta_4
{\tilde\theta}^4 + \tilde\eta_4\partial_s{\tilde\theta}^4 -\tilde \theta_4
\partial_s{\tilde\eta}^4\right)\Big]\nonumber\\&
+ \partial_t \tilde z^M\, \tilde h_M + \frac{4\,i}{\tilde w^3}\, \tilde C\, \left( \partial_s \tilde x 
- \tilde x \right) - \frac{2i}{\tilde w^2}\, \partial_s \tilde z^M\, \tilde\ell_M  
\\\label{eq:Lagrangian_exp_fin}
{\cal L}_F^{(4)} &= \frac{1}{\tilde w^2}\, \tilde B \,.
\end{align}
In the above expression $\tilde B$, $\tilde C$, $\tilde h_M$ and $\tilde \ell_M$ are obtained form the 
quantities $ B$, $  C$, $  h_M$ and $ \ell_M$ in \eqref{eq:pieces} by just replacing the original fields with the corresponding fluctuations.

From the action \eqref{eq:action} we see that the excitations consist of one heavy scalar $\varphi$ of mass $m_{\varphi}^2=4$, one light scalar $x$ with mass $m_{x}^2=2$, six real massless scalars $z^a$ and $\bar z_a$, six massive fermions $\eta^a$ and $\theta^a$ of mass $m_{\psi}^2=1$ and two massless fermions $\theta_4$ and $\eta_4$.
The low-energy asymptotic excitations are the massless scalars coupled to the massless fermion, as in the Bykov model \cite{Bykov:2010tv}.

Compared to \cite{Bianchi:2014ada} we introduced an additional factor of 2 in the redefinition of the worldsheet coordinates to ease the comparison with results from integrability. This effectively doubles the masses of the excitations.
Let us also mention that an alternative form of the Lagrangian can be derived, where the rotation on fermions is undone at the price of introducing a covariant derivative. This Lagrangian in the WZ parametrization is spelled out in appendix \ref{app:lagWZ}.
In order to consider fermionic asymptotic states it could be useful to express the fermionic degrees of freedom in term of Dirac spinors. The Lagrangian is expressed in such a form in appendix \ref{sec:lagdirac}.

\subsection{Feynman rules}

Provided with an explicit Lagrangian for the fluctuations around the cusp background, we can expand it and extract the relevant Feynman rules for performing perturbative computations.

The bosonic propagators are diagonal and read
\begin{equation}
G_{\varphi\varphi}(p) = \frac{1}{T}\, \frac{1}{p^2+4}\qquad\qquad G_{z_a \bar z^b}(p) = \frac{1}{T}\, \frac{2\, \delta_a^b}{p^2}\qquad\qquad G_{xx}(p) = \frac{1}{T}\, \frac{1}{p^2+2}\,.
\end{equation}
The fermionic propagators are not diagonal and, instead, take the form
\begin{align}
 G_{\eta_4\eta^4}(p) &= G_{\theta_4\theta^4}(p) = \frac{1}{T}\, \frac{p_0}{p^2} &
 G_{\eta_4\theta^4}(p) &= G_{\theta_4\eta^4}(-p) = -\frac{1}{T}\, \frac{p_1}{p^2}\nonumber \\
 G_{\eta_a\eta^b}(p) &= G_{\theta_a\theta^b} (p)= \frac{1}{T}\, \frac{p_0}{p^2+1}\delta_a^b &   G_{\eta_a\theta^b}(p) &=G_{\theta_a\eta^b}(-p)= -\frac{1}{T}\, \frac{p_1+i}{p^2+1}\delta_a^b \,.
\end{align}
The interaction vertices are obtained expanding the Lagrangian \eqref{eq:action} in the fluctuation fields. For the one-loop computation only terms with up to four fields are relevant.
They are the same as those of \cite{Bianchi:2014ada} and we spell them out in the appendix \ref{app:lagr_exp}, for completeness.

\section{One-loop dispersion relations}\label{sec:dispersion}

In this section we compute the one-loop corrections to the two-point functions of the elementary fields of the action \eqref{eq:action}.
One-loop self-energy diagrams come in three different topologies: a bubble, a 1PI tadpole and a non-1PI tadpole contributions, which are depicted in Figure \ref{fig:diagrams}.
\FIGURE{
\centering
\includegraphics[width=0.8\textwidth]{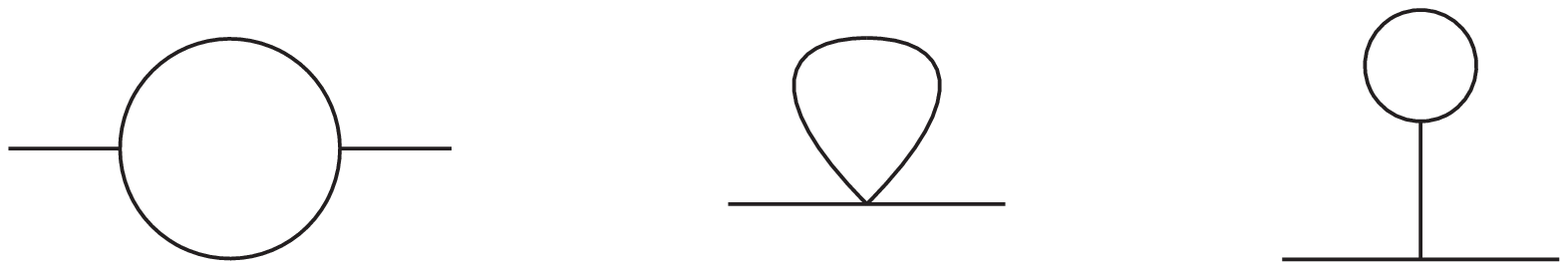}
\caption{Diagram topologies for the two-point function one-loop corrections.}
\label{fig:diagrams}
}
The latter are allowed since the heavy scalar $\varphi$ has a non-trivial expectation value \cite{Bianchi:2014ada}.
Indeed the only one-loop contribution comes from a fermionic loop giving
\begin{equation}
\langle \varphi \rangle = 3\, {\rm I}[1]\,,
\end{equation}
with the tadpole integral ${\rm I}[m^2]$ defined below in \eqref{eq:masterint}.
Bubble and tadpole diagrams give rise to integrals with several powers of loop momentum (up to six) in the numerator.
These are reduced to scalar integrals via Passarino-Veltman reduction. According to the regularization procedure introduced in \cite{Giombi:2009gd}, we perform such a reduction in strictly two dimensions. This entailed consistent results for the computation of the free energy up to two loops both in the $AdS_5\times S^5$ and $AdS_4\times \mathbb{CP}^3$ cases and the dispersion relations in $AdS_5\times S^5$ at one loop. Therefore we expect this choice of regularization to be suitable for the present case as well.

After tensor reduction one is left with two kinds of integral: tadpoles and bubbles
\begin{align}
{\rm I}[m^2] &\equiv \int \frac{d^2k}{(2\pi)^2}\, \frac{1}{k^2+m^2}
\nonumber\\
{\rm I}[m_1^2,m_2^2] &\equiv \int \frac{d^2k}{(2\pi)^2}\, \frac{1}{\left[k^2+m_1^2 \right]\left[(k+p)^2+m_2^2 \right]}\label{eq:masterint}\,.
\end{align}
The latter are ultraviolet convergent and IR finite if both propagators are massive and evaluate to 
\begin{equation}
{\rm I}[m_1^2,m_2^2] = \frac{\log \frac{p^2+m_1^2+m_2^2+\sqrt{(p^2+m_1^2+m_2^2)^2-4\, m_1^2\, m_2^2}}{p^2+m_1^2+m_2^2-\sqrt{(p^2+m_1^2+m_2^2)^2-4\, m_1^2\, m_2^2}}}{4\,\pi\, \sqrt{(p^2+m_1^2+m_2^2)^2-4\, m_1^2\, m_2^2}}\,.
\end{equation}
Whenever one of the masses vanishes the bubble suffers from infrared singularities which can be isolated in terms of tadpole integrals using \cite{Giombi:2010bj}
\begin{equation}\label{eq:bubble0}
{\rm I}[0,m^2] = \frac{1}{p^2+m^2} \left( \frac{1}{2\pi} \log \frac{p^2+m^2}{m^2} - {\rm I}[m^2] + {\rm I}[0] \right)\,.
\end{equation}
Tadpoles are UV divergent. We verify that in dispersion relations they always drop out because they are multiplied by factors going to zero on-shell.
Nevertheless they are present in the off-shell corrections to the two-point functions. In some cases they appear in finite combinations, but in other they do produce ultraviolet singularities, indicating that the corresponding fields undergo a non-trivial wave function renormalization. In particular we observe that the heavy scalar $\varphi$ and the light one $x$ are UV and IR finite even off-shell. On the other hand massless scalars and massive fermions are UV and IR divergent off-shell, though their divergence vanishes on-shell. Curiously massless fermions are UV (and not IR) divergent off-shell, while they are also finite when the on-shell condition is imposed. It is interesting to note how IR divergences appear for particles with a $SU(3)\subset SU(4)$ structure. In $AdS_5\times S^5$ a similar phenomenon was observed for quantities with a $SO(5)\subset SO(6)$ structure and it was expected from the theorems in \cite{Elitzur:1978ww,David:1980gi}, implying that IR divergences appear in any non-$SO(6)$ 
invariant quantities. In the case at hand a similar mechanism is expected to work. Indeed, the $SU(4)$ symmetry of the $\mathbb{CP}^3$ sigma-model is broken to $SU(3)$ by the choice of the vacuum and the appearance of IR divergence are a signal of the non-existence of two-dimensional Goldstone modes. The new feature of the $AdS_4\times \mathbb{CP}^3$ model is the presence of a massless Dirac fermion in the low-energy description \cite{Bykov:2010tv}, which was not present in the higher-dimensional case. As we mentioned, the dispersion relation of this massless Dirac fermion does not contain IR divergences as one should expect being it in the singlet of $SU(4)$.  As far as UV divergences are concerned, their appearance in the off-shell two-point function is not a novel feature for UV-finite string sigma models and recently it was found to be present also in the near-BMN expansion of superstring in $AdS_n\times S^n \times T^{10-2n}$ integrable backgrounds \cite{Roiban:2014cia}.

We collect the tree level structure of propagators according to
\begin{equation}
\langle \bullet(p)\star(-p) \rangle^{(1)} = \frac{1}{T}\, \frac{G_{\bullet\star}(p)}{p^2+m_{\bullet}^2}\, F_{\bullet\star}^{(1)}\,,
\end{equation}
for generic fields $\bullet$ and $\star$. When performing the usual one-loop resummation of non-1PI contributions the on-shell ($p_0=\sqrt{-m^2-p_1^2}$) value of the function $F_{\bullet\star}^{(1)}$ shifts the pole of the propagator. From this shift one can read off the corrections to the dispersion relations in \eqref{eq:result}. In particular evaluating the shift at $p_1=0$ one computes the mass shift $q$ in equation \eqref{eq:result} and subsequently the coefficients $c$ and $d$ by subtraction.
We now spell out the details of the results for the perturbative one-loop corrections to the dispersion relations and masses of each particle in the fluctuation Lagrangian \eqref{eq:action}.

\subsection{Light scalar}

The $x$ scalar self-energy one-loop correction reads
\begin{align}
F_{xx}^{(1)} & = \frac{\left(p_1^2+1\right) \left(-12\, p^2 {\rm I}[1,1] \left(p^4+4\, p_1^2\right)-16 \left(p^4+8\, p^2+4\right) {\rm I}[2,4] \left(p^2-2\, p_1^2\right)\right)}{p^4}+
\nonumber\\&
+ \frac{16 \left({\rm I}[2]-{\rm I}[4]\right) \left(p^2+2\right) \left(p_1^2+1\right) \left(p^2-2 p_1^2\right)}{p^4}\,,
\end{align}
where the difference of UV divergent tadpoles gives a finite remainder ${\rm I}[2]-{\rm I}[4] = \log 2$ and hence $x$ does not need any renormalization.
The self-energy evaluated on-shell reads
\begin{equation}
F_{xx}^{(1)} \Big|_{p^2=-2} = \left(p_1^2+1\right)^2\,.
\end{equation}
The one-loop corrected dispersion relation then becomes
\begin{equation}
p^2+2 = \frac{1}{2\sqrt{2\lambda}}\, F_{xx}^{(1)} \Big|_{p^2=-2} + {\cal O}(\lambda^{-1})\,,
\end{equation}
that is, in Lorentzian signature $(p_0,p_1)\rightarrow (-i E, \ppp)$
\begin{equation}\label{eq:dispx}
E^2 = \ppp^2 + 2 - \frac{1}{4\, h(\lambda)}\, \left(\ppp^2+1\right)^2 + {\cal O}(\lambda^{-1})\,.
\end{equation}
At $\ppp=0$ one can read off the one-loop correction to the mass
\begin{equation}
m_{x}^2 = 2 - \frac{1}{4\, h(\lambda)} + {\cal O}(\lambda^{-1}) < 2\,.
\end{equation}
The fact that the first perturbative correction to the mass at strong coupling is decreasing its value is in general agreement with the trend put forward in \cite{Basso:2010in}, according to which the masses of the gauge excitations should tend to 1 at weak coupling.

\subsection{Heavy scalar}

We now turn to the heavy scalar mode  $\varphi$, whose one-loop correction to the self-energy is found to be
\begin{align}\label{eq:phi2pt}
F_{\varphi\varphi}^{(1)} & = 4 \left(3 {\rm I}[1]-{\rm I}[2] - 2 {\rm I}[4] \right) \left(p^2+4\right) -\frac{12 \left(p^2+4\right) p_1^2 {\rm I}[1,1] \left(p^4+4\, p_1^2\right)}{p^4}
\nonumber\\&
+\frac{8 \left(p^2+4\right)^2 {\rm I}[4,4] \left(p^2-2\, p_1^2\right)^2}{p^4}+2 {\rm I}[2,2] \left(\frac{64\, p_1^4}{p^4}-\frac{64\, p_1^2}{p^2}+\left(p^2+4\right)^2\right)\,.
\end{align}
Again, the difference of UV divergent tadpoles leave a finite remainder $3 {\rm I}[1]-{\rm I}[2] - 2 {\rm I}[4] = 5\log 2$. Therefore the field $\varphi$ does not renormalize, to one loop order.
Evaluating the self-energy on-shell we obtain
\begin{equation}
F_{\varphi\varphi}^{(1)} \Big|_{p^2=-4} =\frac12 p_1^2 \left(p_1^2+4\right)\,.
\end{equation}
In going on-shell the integral ${\rm I}[1,1]$ is singular, which is explained as coinciding with the threshold energy for production of a pair of fermions. This integral is multiplied by a power of $(p^2+4)$, enforcing the limit to vanish.
Then the one-loop corrected dispersion relation reads
\begin{equation}
p^2+4 = \frac{1}{2\sqrt{2\lambda}}\, F_{\varphi\varphi}^{(1)} \Big|_{p^2=-4} + {\cal O}(\lambda^{-1})\,.
\end{equation}
Switching to Lorentzian signature it becomes
\begin{equation}\label{eq:dispphi}
E^2 = \ppp^2 + 4 - \frac{1}{8\, h(\lambda)}\, \ppp^2 \left(\ppp^2+4\right) + {\cal O}(\lambda^{-1})\,.
\end{equation}
The one-loop correction to the mass is clearly seen to vanish. This agrees with the analysis of \cite{Alday:2007mf}, according to which the mass of this mode is protected.
In section \ref{heavy} we discuss more deeply the analytic structure of the one-loop correction \eqref{eq:phi2pt} and its implications for the r\^ole of the heavy scalar in the asymptotic states of the model.

\subsection{Massless scalars}

The one-loop contribution to the two-point function of the massless scalars suffers from both IR and UV divergences, which can be expressed in terms of tadpoles using the identity \eqref{eq:bubble0}.
The $z$ scalar self-energy one-loop correction reads
\begin{align}
F_{zz}^{(1)} & = \frac{1}{2 \pi  p^4} \left[8 \pi  p^2\, {\rm I}[1,1] (p^2-p_1^2) \left(p^4+4\, p_1^2\right)+2 \left(p^2+4\right) \left(p^4-8 p^2 p_1^2+8 p_1^4\right) \log \left(\tfrac{p^2+4}{4}\right) + 
\right.\nonumber\\&\left. 
+ \left(p^6-p^4 \left(2 p_1^2+1\right)+8 p^2 p_1^2-8 p_1^4\right) \log \left(p^2+1\right) \right] + \frac43 \left({\rm I}[0]-3{\rm I}[1]\right) p^2\,.
\end{align}
Then one can see that ${\rm I}[4]$ tadpoles cancel and the rest is proportional to ${\rm I}[0]-3\, {\rm I}[1]$ which is UV (and IR) divergent, but it is multiplied by $p^2$ and vanishes on-shell. 
The on-shell self-energy evaluates
\begin{equation}
F_{zz}^{(1)} \Big|_{p^2=0} = \frac{11}{3 \pi}\, p_1^4\,,
\end{equation}
where the residual UV and IR divergences disappear.
Hence the one-loop corrected dispersion relation reads
\begin{equation}\label{eq:dispz}
E^2 = \ppp^2 - \frac{1}{h(\lambda)}\, \frac{11}{12\, \pi}\, \ppp^4 + {\cal O}(\lambda^{-1})\,.
\end{equation}
At $\ppp=0$ one can read off the one-loop correction to the mass, which is seen to vanish.

\subsection{Massive fermions}

The kinetic terms of the fermion Lagrangian mix the fermion fields.
Hence we have to consider separately the corrections to the two-point functions $\braket{\eta_a \eta^a}$, $\braket{\theta_a \theta^a}$ and $\braket{\eta_a \theta^a}$.
Their computation involves several contributions and the final forms are not particularly illuminating; we spell them out in appendix \ref{app:fermse}.
We point out that the off-shell one-loop corrections to $\langle\eta_a\eta^a\rangle$ and $\langle\theta_a\theta^a\rangle$ are finite, whereas that for $\langle\eta_a\theta^a\rangle$ is UV divergent, although the divergent term cancels on-shell. This implies that the massive fermions, like the massless scalars, undergo wave-function renormalization. The correction to the $\braket{\eta_a \theta^a}$ two-point function is also IR divergent off-shell. Once more the divergent term vanishes on-shell. We will comment on the r\^ole of IR divergences in section \ref{sec:comparison}. 

The different two-point functions all coincide on-shell corroborating the hypothesis that all the massive fermions have the same dispersion relation
\begin{align}
F_{\eta_a\eta^a}^{(1)} \Big|_{p^2=-1} = 
F_{\theta_a\theta^a}^{(1)} \Big|_{p^2=-1} = 
F_{\eta_a\theta^a}^{(1)} \Big|_{p^2=-1} = 2\, p_1^2 \left( p_1^2 + 1 \right)\,.
\end{align}
Thus the one-loop corrected dispersion relation takes the form
\begin{equation}\label{eq:dispferm1}
E^2 = \ppp^2 +1 - \frac{1}{2\, h(\lambda)}\, \ppp^2 \left( \ppp^2 + 1\right)\,,
\end{equation}
from which one sees that the mass does not receive corrections.
Again, this conclusion is in agreement with the integrability prediction that the massive fermion mass is protected from strong to weak coupling.

\subsection{Massless fermions}

The two-point functions for massless fermions are different, depending on the fields, and are UV but not IR divergent. Nevertheless they coincide on-shell, where they are all finite
\begin{equation}
F_{\eta_4\eta^4}^{(1)} \Big|_{p^2=0} = 
F_{\theta_4\theta^4}^{(1)} \Big|_{p^2=0} = 
F_{\eta_4\theta^4}^{(1)} \Big|_{p^2=0} = \frac{p_1^2 \left(7 p_1^2-4\right)}{\pi }\,.
\end{equation}
Hence the one-loop correction to the dispersion relation reads
\begin{equation}\label{eq:dispferm2}
E^2 = \ppp^2 - \frac{1}{4\, \pi\, h(\lambda)}\, \ppp^2 \left( 7 \ppp^2 - 4 \right)\,,
\end{equation}
from which the mass is not corrected.
The massless fermions are a novel feature of the $AdS_4\times \mathbb{CP}^3$ (ABJM) with respect to the $AdS_5\times S^5$ (${\cal N}=4$) case, and we comment on the correction \eqref{eq:dispferm2} to their dispersion relation in the next section.

\section{Comparison with ${\cal N}=4$ SYM and integrability predictions} \label{sec:comparison}

The physics of the excitations on top of the GKP vacuum for the ABJM model has been extensively analysed using integrability in \cite{Basso:2013pxa}. In particular the dispersion relations of its modes were computed exactly. The Bethe ansatz analysis reveals a remarkable similarity with respect to the $AdS_5\times S^5$ spinning string setting.
Therefore we start commenting on the results of the previous section by comparing them with the corresponding findings of ${\cal N}=4$ SYM.
We observe that all the dispersion relations for massive modes are related to those of the corresponding fields in the $AdS_5\times S^5$ sigma model, by
\begin{equation}
E(\ppp)^{(1)}_{AdS_5\times S^5} = E(\ppp)^{(1)}_{AdS_4 \times \mathbb{CP}^3}\,.
\end{equation}
For massless modes such a comparison is not possible, since it is not even clear what to compare: in $AdS_5\times S^5$ there are only massless scalars, whereas for $AdS_4\times \mathbb{CP}^3$ these are coupled to a massless fermion.
Moreover, the scalar excitations over the GKP vacuum in the integrability analysis of \cite{Basso:2013pxa} transform in the $\mathbf{4}$ and $\mathbf{\bar 4}$ of $SU(4)$ whereas the superstring elementary excitations transform only in the fundamental representation of the $SU(3)$ symmetry which survives in the Goldstone vacuum. This is similar to what happens in $\mathcal{N}=4$ SYM where the scalar excitations in the string picture are organized in vectors of $SO(5)$, the explicit symmetry of the $O(6)$ sigma model expanded around the Goldstone vacuum\footnote{In this contest the analysis of \cite{Elitzur:1978ww} gives a recipe for computing $O(N)$ invariant correlation functions in the $O(N)$ sigma model and in \cite{David:1980gi} it was proven that they are free of IR divergences. It is an interesting question whether the same technique can be applied to the Bykov model or even to the full non-linear string sigma model in $AdS_5\times 
S^5$ or in $AdS_4 \times \mathbb{CP}^3$.}.
The dynamics of the massless Dirac fermion is also deeply non-perturbative and can be understood from the low-energy Bykov model~\cite{Bykov:2010tv}.
This is a $\mathbb{CP}^3$ sigma model coupled to a Dirac fermion, with $SU(4)\times U(1)$ symmetry group. Contrary to the bosonic $\mathbb{CP}^3$ sigma model, which exhibits confinement of the scalars, the addition of the fermions makes its rather different, nonperturbatively.
Indeed the fermion forms a chiral condensate which breaks the $U(1)$ symmetry spontaneously and make the gauge field of the $\mathbb{CP}^3$ sigma model massive and dynamical. This in turn prevents it from confining the scalars, which become the spinons of the Bykov model~\cite{Basso:2012bw}.
Hence the massless fermion is not an asymptotic degree of freedom of the theory.

Because of all these differences in the spectra of the superstring and integrability descriptions, as in the ${\cal N}=4$ SYM case, a comparison between their results is only partially possible.
We start commenting on massive modes.
In the asymptotic Bethe ansatz approach the dispersion relation of the massive modes of ${\cal N}=4$ SYM is predicted to be the same as that of the corresponding massive excitations of ABJM.

For the bosons, the quantum correction to the dispersion relation of the light massive scalar agrees with the integrability result.

The heavy scalar, as in ${\cal N}=4$ SYM, is absent in the Bethe ansatz description. Therefore its r\^ole in the sigma model should be analysed carefully and we postpone a thorough discussion of this issue to section \ref{heavy}. Here we just stress that at one loop order the heavy scalar has the same dispersion relation as the corresponding heavy field in ${\cal N}=4$ SYM. 
For fermions, the one-loop corrected dispersion relation for massive modes is in full agreement with the integrability prediction.

Turning to the massless scalar modes, only the fact that the mass does not receive perturbative corrections is compatible with the integrability predictions.
Indeed, the Bethe equations analysis reveals that the model has a gap and such modes acquire an exponentially small mass, non-perturbatively. This parallels what occurs to the scalars of the $O(6)$ sigma model emerging in $AdS_5\times S^5$ in the Alday-Maldacena limit \cite{Alday:2007mf,Basso:2008tx}.
Apart from that, there is no direct identification between the dispersion relations of massless fields of the superstring description and the non-perturbative modes of integrability.
Insisting in comparing the two results at the same order in $\lambda$ just shows that they do not match.
An explanation to this phenomenon, as pointed out in \cite{Zarembo:2011ag}, originates from the presence of perturbatively massless fields. This induces IR divergences in loop computations, which appear as logarithms of the infrared scale of the theory.
Indeed the explicit computation of some one-loop two-point functions already shows the presence of IR divergences, though they always drop out from the dispersion relations.
The infrared cutoff of the theory is set by the non-perturbative mass of the particles which, roughly, scales exponentially with the coupling $\sqrt{\lambda}$. This implies that logarithms of this scale behave like powers of the coupling, effectively lowering the perturbative order to which these terms contribute. In practice this means that an IR divergence appearing at $l$ loops contributes to the $(l-1)$-loop result, invalidating the perturbation theory predictivity at that order.
Therefore it is likely that the one-loop dispersion relations for massless modes \eqref{eq:dispz} and \eqref{eq:dispferm2} are not trustworthy due to two-loop IR divergences despite being IR finite at one loop.
This argument could actually spoil the computation of the one-loop dispersion relations for massive fields, where IR divergences could also appear at two loops. However the theorems in \cite{Elitzur:1978ww,David:1980gi} suggest that $O(6)$ invariant quantities should be IR finite and since $\varphi$ and $x$ are singlets under $O(6)$ we expect their correlation functions to be reliable in perturbation theory.
It would be interesting to ascertain this explicitly via a two-loop computation of the two-point functions.

We finally comment on the massless fermion. This computation should not be plagued by the IR problems which affect massless scalars and hence be reliable.
The result \eqref{eq:dispferm2} indicates that the dispersion relation of the massless fermion acquires a perturbative correction.
In particular, in a small momentum (energy) expansion, we observe a quantum correction to the speed of light (i.e. the coefficient of the $\ppp^2$ term).
This term looks odd, since in this limit the physics is governed effectively by the $\mathbb{CP}^3$ sigma model coupled to a fermion, which is relativistic and does not predict such a correction.
This apparent clash may be reconciled recalling that in the Bykov model nonperturbative effects induce confinement of the fermion which is thus not expected to be present in the full spectrum of excitations at finite coupling (which consists of 4+4 spinons). 
Consequently, as the fermion is not an observable degree of freedom, we do not attribute to the computation of the perturbative correction to its two-point function \eqref{eq:dispferm2} any deep physical meaning.

\section{Comments on the heaviest scalar}\label{heavy}

As is the case for ${\cal N}=4$ SYM, the heaviest scalar mode $\varphi$, which is present in the Lagrangian \eqref{eq:action}, does not correspond to an elementary excitation in the Bethe ansatz description, based on the conjectured integrability of the model.
The r\^ole of this field was deeply analysed in the literature for $AdS_5\times S^5$ \cite{Giombi:2010bj,Zarembo:2011ag,Basso:2014koa,Fioravanti:2015dma}.
A possible explanation that was put forward to explain this mismatch is that the $\varphi$ field is not an asymptotic state of the quantum theory, along the lines of the arguments of \cite{Zarembo:2009au}.
This latter hypothesis and its consequences can be studied perturbatively.
In particular the analytic structure of the two-point function should tell whether it exists as an asymptotic state and whether it is stable or it can decay into lighter particles, such as a pair of massive fermions. This kind of analysis was performed at one loop in \cite{Giombi:2010bj} and \cite{Zarembo:2011ag}.
The punchline is that up to one-loop order the scalar $\varphi$ is a stable threshold composite state of two fermions. Its would be pole in the two-point function coincides with the branching point of the two-fermion continuum square root and hence the scalar cannot be interpreted as a genuine asymptotic bound state.
However, depending on the next order corrections, this conclusion can vary according to how the $\varphi$ and the fermion dispersion relations get modified.

In \cite{Basso:2014koa} the contribution of the heavy scalar appears naturally as a $SU(4)$-singlet compound state of two fermions which perfectly reproduces one of the two-particle contributions to the excited flux-tube. The energy and the momentum of this two-particle state at finite coupling are simply related to the energy and momentum of the fermionic excitations. In particular analysing this relation at strong coupling one finds that
\begin{equation}\label{eq:phifermion}
E_{\varphi}(\ppp)-2\, E_{\psi}\left(\frac{\ppp}{2}\right)=-\frac{\p^2 \ppp^4(\ppp^2+4)^{\frac32}}{8\, \l}+\mathcal{O}(\l^{-\frac32})\,,
\end{equation}
where $\lambda$ is the ${\cal N}=4$ SYM 't Hooft coupling.
The minus sign in the r.h.s of this equation predicts that at two-loops the pole of the heavy scalar two-point function actually moves below the threshold. The results of \cite{Basso:2014koa} show that this property holds also at finite coupling preventing $\varphi$ from decaying into two fermions. Although the pole of the heavy scalar two-point function is shifted below the threshold, the analysis of the singlet channel in the scattering phase of two fermions shows that the unwanted pole is located in the unphysical strip of the rapidity complex plane \cite{Basso:2014koa, Zamolodchikov:2013ama}. This in turn means that $\varphi$ cannot be a true asymptotic state of the theory.

The same arguments should also apply to the heavy scalar in the $AdS_4\times \mathbb{CP}^3$ model.
However they go beyond the one-loop computation carried out in this paper.
What our analysis can test is the integrability prediction that up to one-loop the $\varphi$ scalar should appear as a stable threshold bound state of two fermions.
This expectation can be verified along the lines of \cite{Giombi:2010bj} and \cite{Zarembo:2011ag} as follows.
The one-loop contribution to the denominator of the resummed two-point function has the form
\begin{equation}
F_{\varphi\varphi}^{(1)}(p) = a_0 + a_{1/2} (p^2+4)^{\tfrac12} + \dots\,,
\end{equation}
where all other terms vanish more rapidly in the vicinity of the tree-level mass condition.
In particular we note the presence of the square root $\sqrt{p^2+4}$. Although it is not immediate to see the emergence of this term from \eqref{eq:phi2pt}, it arises from the denominator of ${\rm I}[1,1]$, appearing in the fermion loop diagram.
Close to the threshold, the inverse corrected two-point function
\begin{equation}
G_{\varphi}^{-1}(p) = p^2 + 4 - \frac{1}{2\sqrt{2\lambda}}\, F_{\varphi\varphi}^{(1)}(p) + {\cal O}(\lambda^{-1})
\end{equation}
vanishes at
\begin{equation}
p^2 = -4 + \frac{1}{2\sqrt{2\lambda}}\, a_0 + {\cal O}(\lambda^{-1})\,,
\end{equation}
where here $a_0 = \tfrac12 p_1^2(p_1^2+4)$.
This location lies below the branch cut threshold induced by the square root, meaning that it corresponds to a genuine pole.
From this one would conclude that the $\varphi$ scalar does represent an asymptotic state of the theory.
However this does not take into account that the physical threshold for fermion production is also shifted by quantum corrections.
One can imagine the structure of the resummed two-point function to all orders to have the form (in Lorentz signature)
\begin{equation}\label{eq:inverseprop}
G_{\varphi}^{-1}(\ppp) = -E^2 + 4 E_{\psi_i}^2\left(\frac{\ppp}{2}\right) - \frac{a_{1/2}}{2\sqrt{2\lambda}} \left(-E^2 + 4 E_{\psi_i}^2\left(\frac{\ppp}{2}\right)\right)^{\tfrac12} + \dots\,,
\end{equation}
where $ 4E_{\psi_i}(\ppp/2)=4-\frac{a_0}{2\sqrt{2\lambda}}+\mathcal{O}(\l^{-1})$ is the quantum corrected dispersion relations of the massive fermions.
Its expansion to first order in $\lambda^{-\tfrac12}$ would be in agreement with the perturbative computation \eqref{eq:phi2pt}, although the latter does not guarantee nor hint that \eqref{eq:inverseprop} should hold at higher order.
Assuming this is the case, the would be pole at $E^2 = 4 E_{\psi_i}^2\left(\frac{\ppp}{2}\right)$ coincides with the branching point of the square root. Moreover if the coefficient of the square root $a_{1/2}$ is positive (as the one-loop computation shows it is the case) no other physical poles are present in the two-point function, but only a pole on the second, unphysical, sheet of the square root, located at
\begin{equation}
E^2 = 4\, E_{\psi_i}^2\left(\frac{\ppp}{2}\right) - \frac{a_{1/2}^2}{8\,\lambda} + {\cal O}(\l^{-\frac32})
\end{equation}
where $a_{1/2}$ can be extracted expanding \eqref{eq:phi2pt} near the threshold and reads
\begin{equation}
a_{1/2} = \frac{3\, \ppp^2 (\ppp^2+4)}{4}
\end{equation}
As a result $\varphi$ does not represent a real asymptotic state of the theory.
Insisting on this logic, we can derive a conjectural analogue of \eqref{eq:phifermion}, for the $AdS_4\times \mathbb{CP}^3$ case 
\begin{equation}
E_{\varphi}(\ppp)-2\, E_{\psi_i}\left(\frac{\ppp}{2}\right)=-\frac{9\, \ppp^4(\ppp^2+4)^{\frac32}}{256\, \l}+\mathcal{O}(\l^{-\frac32})\,,
\end{equation}
which would be interesting to check against an integrability based prediction and a full two-loop perturbative computation.

\section{Bound states}\label{sec:boundstates}

The Bethe equation analysis of the GKP excitations shows that the light scalars $x$ can form bound states, whose energy can be computed.
These are not immediately detectable from the superstring approach, however, following \cite{Zarembo:2011ag} we can attempt to estimate their energy to leading order.
This is done treating the $x$ fields as non-relativistic and computing the scattering amplitude of a pair of them.
From this one can extract the effective (attractive) potential experienced by the two particles.
In particular, this is done by computing their $2\to 2$ scattering amplitude and comparing it with the Born approximation in quantum mechanics
\begin{equation}
{\cal M}(k) = - 2\, (2m)^2 \int dx\, e^{-ikx}\, V(x)\,,
\end{equation}
where $k$ is the momentum transfer of the scattering process. This means that the effective potential $V(x)$ is basically the Fourier transform of the amplitude up to numerical constants due to different normalization of the wave-function and Bose statistics.
To lowest order in a momentum expansion, the scattering amplitudes become constants and their Fourier transform is proportional to a $\delta$-function.
The problem then reduces to a many-body system of particles interacting pairwise with a $\delta$-function potential $V_{ij}(x)= - g\, \delta(x_{i}-x_{j})$.
Such a model admits a two-particle bound state with one energy level $E=-\tfrac{\mu\,g^2}{2}$, where $\mu$ is the reduced mass of the system ($\mu=\tfrac{1}{\sqrt{2}}$ for the $x$ scalars).
More generally the binding energies for bound states of $\ell$ particles of mass $m$ are also known \cite{McGuire:1964zt}
\begin{equation}\label{eq:multidelta}
E_{\ell} = - \frac{m\, g^2}{24}\, \ell (\ell^2-1)\,.
\end{equation}
This energy can be compared to the static limit of the lowest order expansion for $\lambda\gg 1$ of the binding energy derived from integrability. This is given by
\begin{equation}
E_{binding,\ell}(\ppp) = E_{\ell}(\ppp) - \ell\, E_1(\tfrac{\ppp}{\ell})\,,
\end{equation}
where $E_{\ell}(\ppp)$ is the dispersion relation for the relevant twist $\ell$ excitation.

In ${\cal N}=4$ SYM such a program was successfully carried out for the gauge excitation, showing agreement with the integrability prediction at $\ppp=0$.
In this section we perform a similar computation for the mass $\sqrt{2}$ mode of the $AdS_4\times \mathbb{CP}^3$ superstring.
At tree level the amplitude for $xx\to xx$ scattering receives contributions from all $s$, $t$ and $u$ channels, as in Figure \ref{fig:scatx}.
\FIGURE{
\centering
\includegraphics[width=0.7\textwidth]{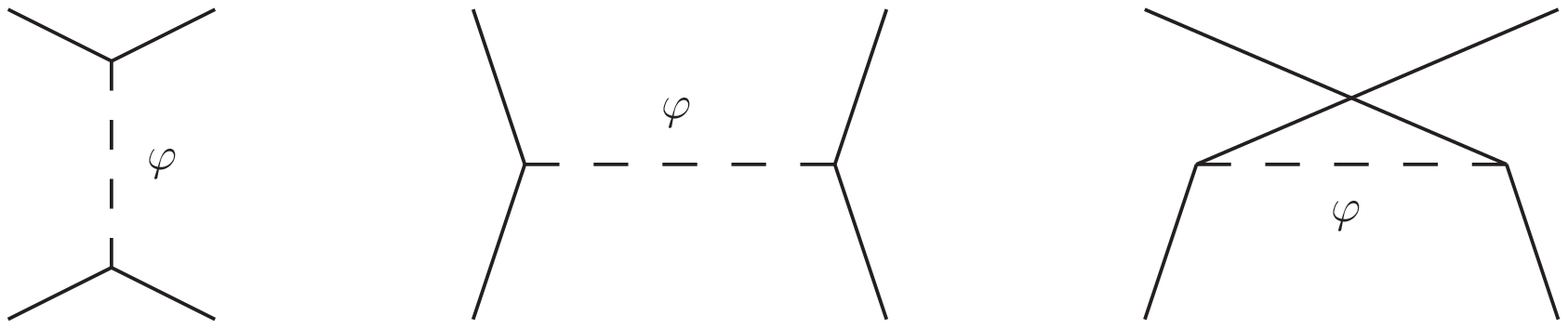}
\caption{Tree level scattering $xx\to xx$.}
\label{fig:scatx}
}
In the zero momenta limit the contributions from the $t$ and $u$ channels are equal and give
\begin{equation}
{\cal M}_{xx\to xx,\, t} = {\cal M}_{xx\to xx,\, u} = 2^5\, \sqrt{2\lambda} + {\cal O}(k)\,,
\end{equation}
whereas the $s$-channel contributes with an opposite result, corresponding to a repulsive interaction.
Altogether the amplitude gives
\begin{equation}
{\cal M}_{xx\to xx} = 2^5\, \sqrt{2\lambda} + {\cal O}(k)\,,
\end{equation}
from which we find the effective potential (after properly rescaling fields by a $T^{-1/2}$ factor and introducing $h(\lambda)$ \eqref{eq:h})
\begin{equation}
V_{xx}(x) = -\frac{1}{4\, h(\lambda)}\, \delta(x)\,.
\end{equation}
Plugging this into \eqref{eq:multidelta} we give an estimate for the binding energy of the twist $\ell$ gauge bound state
\begin{equation}
E_{binding,\ell}(0) = - \frac{\sqrt{2}\, \ell (\ell^2-1)}{384\, h(\lambda)^2} + {\cal O}(\lambda^{-2})\,,
\end{equation}
which is equivalent to the corresponding one for $AdS_5\times S^5$, once the replacement $h(\lambda) \to \tfrac{\sqrt{\lambda}}{4\pi}$ is performed. Thus it agrees with the integrability prediction of \cite{Basso:2010in} at first order at strong coupling.

According to the parallel analysis of \cite{Basso:2014koa} in $AdS_5\times S^5$, multi-fermion states are also present in the theory.
These appear as bound states of the two-fermion composites which we have identified as the mass 2 excitations $\varphi$ of the sigma model. This composite states of $2n$ fermions are expected to have mass $2n$ and consequently the bound states of $\varphi$ to have zero binding energy at vanishing momentum\footnote{We would like to thank B. Basso for explaining this to us.}.
We therefore repeat the same analysis as above for the scalars $\varphi$, in order to check whether the binding energy is vanishing at leading order in the static limit.
\FIGURE{
\centering
\includegraphics[width=0.9\textwidth]{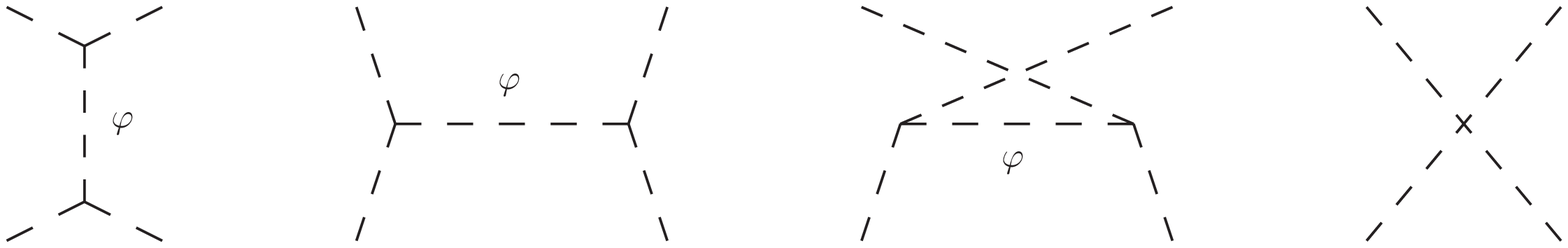}
\caption{Tree level scattering $\varphi\varphi \to \varphi\varphi$.}
\label{fig:scatphi}
}
The lowest order scattering amplitude for $\varphi\varphi\to \varphi\varphi$ is given by the sum of the diagrams in Figure \ref{fig:scatphi}. Once again the $t-$ and the $u-$channel give two identical contributions in the static limit
\begin{equation}
 {\cal M}_{\varphi\varphi\to \varphi\varphi,\, t} = {\cal M}_{\varphi\varphi\to \varphi\varphi,\, u} = 2^7\, \sqrt{2\lambda} + {\cal O}(k)\,.
\end{equation}
In this case also the four point vertex gives an attractive contribution which is once more equal to
\begin{equation}
 {\cal M}_{\varphi\varphi\to \varphi\varphi,\, 4}= 2^7\, \sqrt{2\lambda} + {\cal O}(k)\,.
\end{equation}
The s-channel contribution, as in the previous case, contributes with a repulsive interaction which compensates exactly the other terms
\begin{equation}
{\cal M}_{\varphi\varphi\to \varphi\varphi,\, s} =-3\times  2^7\, \sqrt{2\lambda} + {\cal O}(k)\,.
\end{equation}
In conclusion
\begin{equation}
{\cal M}_{\varphi\varphi\to \varphi\varphi} =  {\cal O}(k)\,,
\end{equation}
which implies that the bound state of $\varphi$ has vanishing binding energy in the static limit in agreement with the integrability prediction.
As a further check we performed the same computation in $AdS_5\times S^5$ where the vertices are modified by relative factors and we found that the mechanism is exactly the same. Therefore, as expected, the binding energy vanishes also in that case.

\section{Conclusions}\label{sec:conclusions}

In this paper we have studied perturbatively the dynamics of the excitations on top of the  GKP vacuum in the $AdS_4\times \mathbb{CP}^3$ string background, which is dual to the ABJM theory, at strong coupling.
This model is conjectured to be integrable, which allowed to solve exactly for several physical quantities such as the dispersion relations of the elementary excitations and their scattering amplitudes.
We have performed a direct perturbative computation of the dispersion relations for the excitations appearing in the superstring description to one loop order at strong coupling.
We have used the light-cone gauge Lagrangian of the $AdS_4\times \mathbb{CP}^3$ sigma model expanded about the cusp background, which is equivalent to the spinning string solution and is more efficient for perturbative computations.

Summarizing our findings:
\begin{itemize}
\item We have ascertained that the dispersion relation for massive modes coincides with that predicted by the asymptotic Bethe ansatz \cite{Basso:2013pxa}. This constitutes a test of integrability of the model, in the strong coupling regime.
\item For massless modes a comparison between the supertring perturbative approach and the integrability modes is not straightforward, as already stressed in the similar context of the $AdS_5\times S^5$ superstring \cite{Zarembo:2011ag}. We verify that this is also true for the ABJM dual sigma model, where it is hard to match the elementary massless modes with the spinons of the integrability description, and consequently there is no clear identification of their dispersion relations. 
\item We have analysed in detail the dispersion relation of the heaviest scalar in the spectrum of the string fluctuations. Such a mode is absent as an elementary state in the integrability approach and in $AdS_5\times S^5$ has been interpreted as a two-fermion state appearing as a pole in the fermion S-matrix, but in the unphysical strip in the rapidity plane, preventing it from being associated to an asymptotic state \cite{Basso:2014koa}. 
Such a pole is expected to appear at two-loop order since up to one loop it merges with the branching point of the two-fermion continuum.
In agreement with the expectation that the same phenomenon could happen in the $AdS_4\times \mathbb{CP}^3$ case, we find that to one loop order the dispersion relation of the heaviest scalar exhibits a pole which coincides with the threshold for production of two fermions.
\item We have estimated the binding energy of bound states of gauge and two-fermion excitations from the non-relativistic limit of their $2\to 2$ scattering amplitude. 
The results are consistent with the integrability predictions in the static approximation.
\end{itemize}

\section*{Acknowledgements}

We thank Benjamin Basso, Valentina Forini, Ben Hoare, Matias Leoni and Stefan Zieme for very useful discussions. We are also particularly grateful to Konstantin Zarembo and Stefan Zieme for sharing their work with us, prior to publication.
The work of LB is funded by DFG via the Emmy Noether Program
``Gauge Fields from Strings''. 
The work of MB was supported in part by the Science and Technology Facilities Council Consolidated Grant ST/L000415/1 \emph{String theory, gauge theory \& duality}.

\vfill
\newpage

\appendix

%%%%%%%%%%%%%%%%%%%%%%%%%%%%%%%%%
\section{Details on the expanded Lagrangian}\label{expansion}
\label{app:lagr_exp}
%%%%%%%%%%%%%%%%%%%%%%%%%%%%%%%%%
In this appendix we provide the expanded fluctuation Lagrangian \eqref{eq:action} up to quartic order in the fields. 
The vertices come with a factor $\frac12$, with respect to the original Lagrangian, from the prefactor $\tfrac{T}{2}$ in the action.
In order not to clutter the expressions we drop the tildes and the coupling $T$, which is understood to appear in each vertex insertion in Feynman diagrams. 
The cubic interactions read
\begin{equation*}
V_{\varphi x x}=-4\varphi \left[(\pa_s-1)\, x\right]^2 \qquad V_{\varphi^3}=2\varphi\left[(\pa_t \varphi)^2-(\pa_s\varphi)^2\right] \qquad V_{\varphi |z|^2}=2\varphi\left[|\pa_t z|^2-|\pa_s z|^2\right]
\end{equation*}
\begin{eqnarray}
\begin{array}{lll}
V_{z\eta\eta}=-\e^{abc} \pa_t \bar z_a \eta_b \eta_c + h.c. &&
 V_{z\eta\theta}=-2\, \e^{abc} \bar z_a \eta_b (\pa_s-1)\theta_c-h.c. \nonumber\\
 V_{\varphi\eta \theta}=-4\, i\, \varphi\, \eta_a (\pa_s-1)\theta^a -h.c. &&
 V_{x \eta \eta}=-4\, i\,  \eta^a \eta_a (\pa_s-1) x \nonumber\\
 V_{z\eta_a\eta_4}=\ -2\, \pa_t z^a \eta_a \eta_4 + h.c. &&
 V_{z\eta_a\theta_4}=2\, \pa_s z^a \eta_a \theta_4-h.c. \nonumber\\
 V_{\varphi \eta_4 \theta^4}= -2\,i\, \varphi\, (\theta^4 \pa_s \eta_4-\pa_s\theta^4 \eta_4)-h.c. &&
 V_{x \psi^4 \psi_4}= -2\, i\, (\eta^4 \eta_4+\theta^4 \theta_4) (\pa_s-1) x 
\end{array}
\end{eqnarray}
whereas the quartic vertices are
\begin{align}
V_{z^4}&=\frac16 \left[ \left(\bar z_a \pa_t z^a\right)^2 + \left(\bar z_a \pa_s z^a\right)^2 + \left(z^a \pa_t \bar z_a\right)^2 + \left(z^a \pa_s \bar z_a\right)^2
\right. & &\nonumber\\& \hspace{1cm} \left.
- |z|^2\left(|\pa_t z|^2+|\pa_s z|^2\right) - \left|\bar z_a \pa_t z^a\right|^2 - \left|\bar z_a \pa_s z^a\right|^2\right]
\end{align}
\begin{eqnarray}
\begin{array}{lll}
V_{\varphi^2 x x} =16\,\varphi^2\, \left[(\pa_s-1)\, x\right]^2 &&
V_{\varphi^4} =4\, \varphi^2\left[(\pa_t \varphi)^2+(\pa_s\varphi)^2+\frac23\varphi^2\right]\\
V_{\varphi^2 |z|^2} =4\, \varphi^2 \left[|\pa_t z|^2+|\pa_s z|^2\right]&&
V_{\dot z\bar z \psi^4 \psi_4} = -2\, i\, (\eta^4 \eta_4+\theta^4 \theta_4)\bar z _b\pa_t z^b+h.c.\\
V_{\eta^2 \eta^4 \eta_4} = 8\,   \eta^4 \eta_4  \eta^a \eta_a  &&
 V_{ z'\bar z \psi^4 \psi_4} = -2\, i\, (\eta^4 \theta_4-\theta^4 \eta_4)\bar z _b\pa_s z^b-h.c.\\
V_{\eta^4} =4 (\eta^a \eta_a)^2&&
 V_{\varphi^2 \eta_4 \theta^4} = 4\,i\, \varphi^2\, (\theta^4 \pa_s \eta_4-\pa_s\theta^4 \eta_4)-h.c.\\
V_{\eta_4 \eta^4\theta_4 \theta^4} =-8\,  \eta^4 \eta_4  \theta^4 \theta_4&&
 V_{\varphi\, x \psi^4 \psi_4} =12\,i\, \varphi\, (\eta^4 \eta_4+\theta^4 \theta_4) (\pa_s-1) x\\
V_{\eta^3 \eta_4} =4\,  \e^{abc}\eta_a\eta_b\eta_c\eta_4+h.c.&&
V_{zz\eta^a\eta_4} =-2\, i\, \e_{abc} \pa_t z^a z^b  \eta^c \eta_4+ h.c.\\
V_{\varphi\, z\eta_a\theta_4} =-8\, \varphi\,  \pa_s z^a \eta_a \theta_4-h.c. &&
 V_{\varphi\, z\eta\theta} =8\, \varphi \e^{abc} \bar z_a \eta_b (\pa_s-1)\theta_c-h.c. \\
V_{zz\eta^a\theta_4} =2\, i\, \e_{abc} \pa_s z^a z^b  \eta^c \theta_4- h.c.&&
  V_{zz\eta\eta} =-2\, i\, (\bar z_a \pa_t z^a \eta^b \eta_b-\bar z_b \pa_t z^a  \eta^b \eta_a) +h.c. \\
V_{\varphi\, x \eta \eta} =24\, i\,\varphi\,  \eta^a \eta_a (\pa_s-1) x &&
  V_{zz\eta\theta} =-2\, i\, [|z|^2 \eta_a (\pa_s-1)\theta^a - \bar z_b z^a\eta_a (\pa_s-1)\theta^b] -h.c. \\
V_{\varphi^2 \eta \theta} =8\, i\, \varphi^2\, \eta_a (\pa_s-1)\theta^a -h.c. &&
 V_{x z\eta\eta} =-4\, (\pa_s-1)x \e^{abc} \bar z_a \eta_b \eta_c - h.c. 
\end{array}
\end{eqnarray} 

\section{Self-energies of fermions}\label{app:fermse}

In this appendix we collect the off-shell fermion self-energies.
For massive fermions they read
\begin{align}
F_{\eta_a\eta^a}^{(1)} &= \frac{2}{p^6} \bigg[
\left(p^2+1\right) \Big(\left(-2 {\rm I}[1]-{\rm I}[2]-{\rm I}[4
]\right) p^6 + 
\nonumber\\&
+p^4 \left(\left(6 {\rm I}[1]+{\rm I}[2]-7 {\rm I}[4]\right) p_1^2-2 {\rm I}[1]+{\rm I}[2]+{\rm I}[4]\right)+
\nonumber\\&
+ p^2 \left( \left(26 {\rm I}[1]-5 {\rm I}[2]-21 {\rm I}[4]\right) p_1^2+ \left(16 {\rm I}[4]-16 {\rm I}[1]\right)p_1^4\right)-4 \left(10 {\rm I}[1]-{\rm I}[2]-9 {\rm I}[4]\right) p_1^4\Big) + 
\nonumber\\&
- \left(p^2+1\right) \Big(
\frac{2 p_1^2\left(p^4-p^2+4\, p_1^2\right) \log \left(p^2+1\right)}{\pi } + 
\nonumber\\&
-\left(3 p^4 + 4 p^6 + p^8 - 63 p^2 p_1^2 - 56 p^4 p_1^2 - 9 p^6 p_1^2 + 
 108 p_1^4 + 108 p^2 p_1^4 + 16 p^4 p_1^4\right) {\rm I}[1,4] \Big) + 
\nonumber\\&
+\left(p^2-p_1^2\right) \left(p^2 \left(p^2+1\right)^3+4 \left(p^4-4\, p^2-1\right) p_1^2\right) {\rm I}[1,2]
\bigg]
\end{align}
\begin{align}
F_{\theta\bar\theta}^{(1)} &= 2 \frac{p_1^2+1}{p^6} \bigg[
\left(p^2+1\right) \Big( 
p^4 \left(6 {\rm I}[1]+{\rm I}[2]-7 {\rm I}[4]\right) +
\nonumber\\&
+ p^2 \left( 10 {\rm I}[1]- {\rm I}[2]-9 {\rm I}[4] + \left(16 {\rm I}[4]-16 {\rm I}[1]\right)p_1^2\right)-4 \left(10 {\rm I}[1]-{\rm I}[2]-9 {\rm I}[4]\right) p_1^2\Big) 
\nonumber\\&
- \left(p^2+1\right) \Big(
\frac{2 \left(p^4-p^2+4\, p_1^2\right) \log \left(p^2+1\right)}{\pi }  
\nonumber\\&
-\left(27 p^2 + 36 p^4 + 9 p^6 - 108 p_1^2 - 108 p^2 p_1^2 - 16 p^4 p_1^2\right) {\rm I}[1,4]\Big)
\nonumber\\&
- \left(p^2 \left(p^2+1\right)^3+4 \left(p^4-4\, p^2-1\right) p_1^2\right) {\rm I}[1,2]
\bigg]
\end{align}
\begin{align}
F_{\eta\bar\theta}^{(1)} &= \frac{2}{p^6} \bigg[
\left(p^2+1\right) \Big(
\left(-4 {\rm I}[0]+2 {\rm I}[1]-{\rm I}[2]+{\rm I}[4]\right) p^6+
\nonumber\\&
+p^4 \left(\left(14 {\rm I}[1]+ {\rm I}[2]-15 {\rm I}[4]\right) p_1^2-4 {\rm I}[1]+{\rm I}[2]+3 {\rm I}[4]\right)+
\nonumber\\&
+p^2 p_1^2 \left(\left(16 {\rm I}[4]-16 {\rm I}[1]\right) p_1^4+38 {\rm I}[1]-5 {\rm I}[2]-33 {\rm I}[4]\right)+4 \left({\rm I}[2]+9 {\rm I}[4]-10 {\rm I}[1]\right) p_1^4 \Big)+ 
\nonumber\\&
- \left(p^2+1\right) \Big(
\frac{2 \left(p^4-3p^2+4\, p_1^2\right) \log \left(p^2+1\right)}{\pi }
\nonumber\\&
+\left(9 p^4 + 12 p^6 + 3 p^8 - 99 p^2 p_1^2 - 100 p^4 p_1^2 - 17 p^6 p_1^2 + 
 108 p_1^4 + 108 p^2 p_1^4 + 16 p^4 p_1^4\right) {\rm I}[1,4]\Big) 
\nonumber\\&
+ \left(p^2-p_1^2\right) \left(p^2(p^2+1)^3+4 \left(p^4-4p^2-1\right) p_1^2\right) {\rm I}[1,2]
\bigg]
\end{align}
For massless fermions they are
\begin{align}
F_{\eta_4\bar\eta^4}^{(1)} &= F_{\theta_4\bar\theta^4}^{(1)} = \frac{1}{4 \pi  p^6}\bigg(6 \left(1 + p^2\right) \left(p^6 + 12 p^2 p_1^2 - 16 p_1^4 - p^4 \left(1 + 4 p_1^2\right)\right) \log \left(p^2+1\right)+
\nonumber\\&
+\left(p^4 \left(4 + p^2\right)^2 - 32 p^2 \left(6 + 5 p^2 + p^4\right) p_1^2 + 
  64 \left(2 + p^2\right)^2 p_1^4\right) \log \left(\tfrac{p^2}{4}+1\right)+
\nonumber\\&
+\left(4 p^4 + p^8 - 48 p^2 p_1^2 + 64 p_1^4\right) \log \left(\tfrac{p^2}{2}+1\right)\bigg)-\left(6 {\rm I}[1]+{\rm I}[2]+{\rm I}[4]\right) p^2
\end{align}
\begin{align}
F_{\eta_4\bar\theta^4}^{(1)} &= \frac{1}{4 \pi  p^6}\bigg(6 \left(1 + p^2\right) \left(3p^6 + 20 p^2 p_1^2 - 16 p_1^4 - p^4 \left(5 + 4 p_1^2\right)\right) \log \left(p^2+1\right)+
\nonumber\\&
+\left(p^4 \left(4 + p^2\right)(20+9p^2) - 32 p^2\left(2 + p^2\right) \left(5 +2 p^2\right) p_1^2 + 
  64 \left(2 + p^2\right)^2 p_1^4\right) \log \left(\tfrac{p^2}{4}+1\right)+
\nonumber\\&
+\left(20 p^4 + p^8 - 80 p^2 p_1^2 + 64 p_1^4\right) \log \left(\tfrac{p^2}{2}+1\right)\bigg)-\left(6 {\rm I}[1]+{\rm I}[2]+{\rm I}[4]\right) p^2
\end{align}

\section{Lagrangian in Wess-Zumino type parametrization}\label{app:lagWZ}
The Lagrangian in Wess-Zumino (WZ) type parametrization was written down in appendix A of \cite{Bianchi:2014ada} and we refer the reader to that reference for further details on the explicit expression. Here we start from that expression\footnote{With respect to \cite{Bianchi:2014ada} we reabsorb a factor of $i$ in the definition of the covariant derivative eliminating it from the definition of ${\Omega_{\hat a}}^{\hat b}$\, .} and we expand it around the null-cusp solution along the same lines of section \ref{sec:expansion}. The fluctuations fields are introduced in \eqref{eq:fluctuations} and plugging them into the lagrangian in WZ type parametrization we find a bosonic Lagrangian which is identical to \eqref{boslag} and a fermionic part given by
\begin{align}
 L^{(2)}_F&=-i\tilde\eta^a\pa_t\tilde\eta_a-i\tilde\eta_a\pa_t\tilde\eta^a-i\tilde\theta^a\pa_t\tilde\theta_a-i\tilde\theta_a\pa_t\tilde\theta^a-\frac{2\, i}{\tilde w^2}[\tilde\eta^a(\pa_s-1)\tilde\theta_a-\tilde\eta_a(\pa_s-1)\tilde\theta^a]\nonumber\\
 &-i\tilde\eta^4\pa_t\tilde\eta_4-i\tilde\eta_4\pa_t\tilde\eta^4-i\tilde\theta^4\pa_t\tilde\theta_4-i\tilde\theta_4\pa_t\tilde\theta^4-\frac{i}{\tilde w^2}(\tilde\eta^4\pa_s\tilde\theta_4-\tilde\eta_4\pa_s\tilde\theta^4-\tilde\theta^4\pa_s\tilde\eta_4+\tilde\theta_4\pa_s\tilde\eta^4)\nonumber\\
 &-\frac{4\, i}{\tilde w^3}(2\tilde\eta^b \tilde\eta_b+\tilde \eta^4 \tilde \eta_4+\tilde \theta^4\tilde\theta_4)(\pa_s-1)\tilde x-2\, {\O_t}^c\tilde\eta^a\tilde \eta^b \e_{acb}+2\, {\O_t}_c \tilde \eta_a \tilde \eta_b \e^{acb}\nonumber \\
 &-\tilde\eta^{\hat a} {{\Omega_t}_{\hat a}}^{\hat b} \tilde\eta_{\hat b}-\tilde\theta^{\hat a} {{\Omega_t}_{\hat a}}^{\hat b} \tilde\theta_{\hat b}-\frac{2}{\tilde w^2} \tilde\eta^{\hat a} {C_{\hat a}}^{\hat b} {{\Omega_s}_{\hat b}}^{\hat c} \tilde\theta_{\hat c}-4\, {{\Omega_t}_a}^a(\tilde\eta^4\tilde\eta_4+\tilde\theta^4\tilde\theta_4)-\frac{4}{\tilde w^2} {{\Omega_s}_a}^a (\tilde\eta^4\tilde\theta_4-\tilde\theta^4\tilde\eta_4)\nonumber\\
 &-4\, \tilde \eta_a \O^a_t \tilde\eta_4+4\, \tilde\eta^a {\O_t}_a \tilde\eta^4+\frac{4}{\tilde w^2}(\tilde \eta_a {\O_s}^a \tilde\theta_4+\tilde \eta^a {\O_s}_a \tilde \theta^4)  \label{eq:lagWZ2}
\end{align}
where we used the short-hand notation 
\begin{align}
 \eta_{\hat{a}}&=\left(\begin{array}{c}
                      \tilde \eta_a \\
                      \tilde \eta^a 
                      \end{array}\right)\, , &  \eta^{\hat{a}}&=({\tilde \eta}^a,\tilde\eta_a)\, ,\\
 {\Omega_{\hat a}}^{\hat{b}}&=\left(\begin{array}{cc}
                              \Omega_{a}^{\phantom{a}b} - \d_a^b \Omega_c^{\phantom{c}c}& \e_{acb}\Omega^c\\
                              -\e^{acb}\Omega_{c} & -\Omega_b^{\phantom{b}a}+\d_b^a \Omega_c^{\phantom{c}c}
                             \end{array}\right)\,, &  {C_{\hat a}}^{\hat b}&=\left(\begin{array}{cc}
                              \d_a^b & 0 \\
                              0 & -\d^a_b
                             \end{array}\right).
\end{align}
The vielbein $\O_a$ and $\O^a$ are given in \eqref{eq:forms}, whereas the spin connection ${\O_a}^b$ reads 
\begin{align}\label{eq:spinconn}
\Omega_{a}^{\phantom{a}b}&=i\frac{(1-\cos{|z|})}{|z|^2}(\bar z_adz^b-d\bar z_az^b)-i\bar z_az^b\frac{(1-\cos{|z|})^2}{2|z|^4}(dz^c\bar z_c-z^cd\bar z_c).
\end{align}

One can check that, after the assignments \eqref{eq:diracspinor}, the lagrangian \eqref{eq:lagdirac2} exactly reproduces \eqref{eq:lagWZ2}.

The expansion can be carried out in the same way for the quartic part which reads
\begin{equation}
 L^{(4)}_F= \frac{8}{\tilde w^2}[(\tilde \eta_a\tilde \eta^a)^2+\varepsilon^{abc}
\tilde \eta_a\tilde \eta_b\tilde \eta_c\tilde \eta_4+\varepsilon_{abc}
\tilde \eta^a\tilde \eta^b\tilde \eta^c\tilde \eta^4+2\,
\eta_4\tilde{\eta}^4\tilde \eta_a\tilde {\eta}^a-2\,\tilde \eta_4 \tilde \eta^4 \tilde \theta_4\tilde \theta^4] \,.
\end{equation}

\section{Lagrangian with Dirac fermions}\label{sec:lagdirac}
When considering the S-matrix for excitations on top of the GKP vacuum it could be useful to rewrite the Lagrangian in an equivalent form where the asymptotic degrees of freedom are more manifest.
Here we start from the lagrangian in the Wess-Zumino type parametrization expanded around the null-cusp solution as shown in appendix \ref{app:lagWZ}. We introduce four Dirac fermions $\{\Psi^{a},\Psi^4\}$ which are related to the previous degrees of freedom by\footnote{Here the fields in the Lagrangian are the same as in \eqref{eq:Lagrangian_exp_fin} although the tilde is omitted for simplicity.}
\begin{equation}\label{eq:diracspinor}
 \Psi^a=\left(\begin{array}{c} \eta^a \\ -\theta^a\end{array} \right) \qquad \Psi^4=\left(\begin{array}{c} \eta^4 \\ -\theta^4\end{array} \right) .
\end{equation}
Following \cite{Zarembo:2011ag} we use the Euclidean gamma matrices 
\begin{equation}
 \gamma^t=\left(\begin{array}{cc} 
                0& -1\\
                -1& 0
                \end{array}\right)\qquad
 \gamma^s=\left(\begin{array}{cc} 
                 1& 0\\
                 0& -1
                \end{array}\right)
\end{equation}
and the projectors $\Pi_\pm=\frac{\mathbb{1}\pm\g^s}{2}$. In order to simplify the expression of the Lagrangian we notice that there is a recurrent structure for the operators in Dirac space which is a generalization of the standard slashed notation reminiscent of the non-conformally flat worldsheet metric. Therefore, given an operator $O$ in Dirac space we introduce the notation $\slashed{O} =O_t \g^t+\frac{1}{\tilde{w}^2} O_s \g^s$. The bosonic part of the Lagrangian is identical to the one given in \eqref{boslag}, whereas the quadratic part in fermions is significantly modified and reads
\begin{align}
 L_F^{(2)}&=-2\, i\, \bar\Psi_a \left(\backslashedpa{\partial}+{\frac{1}{w^2}}\right)\Psi^a-2\, i\, \bar\Psi_4 \backslashedpa{\partial}\Psi^4-2\, i\, \bar \Psi_a \Pi_+ \Psi^a \pa_s {\frac{1}{w^2}}-i\, \bar\Psi_4 \g^s \Psi^4 \pa_s {\frac{1}{w^2}}\nonumber\\
 &+{\frac{4\, i}{w^3}}\left(2\, \bar\Psi_a \g^t \Pi_+ \Psi^a+\bar \Psi_4 \g^t \Psi^4\right)\left(\pa_s-1\right)x+2\, \bar\Psi_a {\slashed{\Omega}_b}^a\Psi^b+2\, \bar\Psi_a {\slashed{\Omega}_b}^b\Psi^a \nonumber\\
 &-\e_{abc}{\Psi^T}^a{\Omega_t}^b(1+2\, \Pi_+)\Psi^c-\frac{1}{w^2}\e_{abc}{\Psi^T}^a\g^t{\Omega_s}^b\Psi^c\nonumber\\
 &+\e^{abc}{\bar\Psi}_a {\Omega_t}_b(1+2\, \Pi_-){\bar\Psi^T}_c-\frac1{w^2}\e^{abc}{\bar\Psi}_a\g^t {\Omega_s}_b{\bar\Psi^T}_c\nonumber\\ 
 &+4\, {\Psi^T}^a \Pi_+ \slashed{\Omega}_a \g^t \Psi^4-4\, {\bar\Psi}_a \Pi_- \slashed{\Omega}^a \g^t {\bar\Psi^T}_4+4\, \bar \Psi_4 {\slashed{\Omega}_a}^a \Psi^4  \label{eq:lagdirac2}
\end{align}
The vielbein $\O_a$ and $\O^a$ and the spin connection ${\Omega_a}^b$ are given in \eqref{eq:forms} and \eqref{eq:spinconn}.
Their time and space components are defined in the obvious way as $\O={\O}_t dt+ {\O}_s ds$ which holds for all the possible indices configuration. Let us stress that the Lagrangian now depends only on the operators $\O_a$, $\O^a$ and ${\O^a}_b$ and all the dependence on the matrices $T$ has been reabsorbed in those quantities. The fact that $\Omega^a$ is expanded only in odd powers of $z$ and ${\Omega_a}^b$ only in even powers starting from $\mathcal{O}(z^2)$ drastically simplifies the expansion of the Lagrangian.
 
The part containing four powers of fermions can be written as 
\begin{multline}
 L_F^{(4)}=+\frac{8}{w^2}[(\bar\Psi_a\g^t\Pi_+ \Psi^a+\bar \Psi_4 \g^t \Pi_+ \Psi^4)^2-(\bar\Psi_4\g^t\Psi^4)^2\\+\e_{abc}{\Psi^T}^a \Pi_+ \Psi^b\ {\Psi^T}^c\Pi_+ \Psi^4+\e^{abc}{\bar\Psi}_a \Pi_- {\bar\Psi^T}_b\ {\bar\Psi}_c\Pi_- {\bar\Psi^T}_4]
\end{multline}

The Lagrangian can be easily expanded up to quartic order in the fields. From the quadratic Lagrangian we can read out the spectrum of the theory
\begin{equation}
L_2=\pa_i x \pa^i x +2\, x^2+ \pa_i \phi \pa^i \phi +4\, \phi^2+\pa_i z^a \pa^i \bar z_a-2\, i\, \bar\Psi_a \left(\slashed{\pa} +1\right)\Psi^a- 2\, i\, \bar\Psi_4\slashed{\pa} \Psi^4\, .
\end{equation}
Here worldsheet indices are raised and lowered with the euclidean two-dimensional flat metric and the slash simply denotes the combination $\slashed{\pa}=\g^s\pa_s+\g^t \pa_t$. As expected the field content of the theory consists of eight bosons, one of mass 4, one of mass $2$ and six massless, and four dirac fermions, three of mass $1$ and one massless. The cubic Lagrangian introduces the interaction and breaks the worldsheet Lorentz invariance yielding a non-relativistic dispersion relation. With Dirac fermions it reads
\begin{align}
 L_3&=4\,\phi\left[-2(\nabla_s x)^2+(\pa_t \phi)^2-(\pa_s \phi)^2+|\pa_t z|^2-|\pa_s z|^2\right.\nonumber\\ 
 &\phantom{=4\, \phi\left[\right.}\left.-2\, i\, \nabla_s \bar\Psi_a \Pi_- \Psi^a+2\, i\, \bar\Psi_a \Pi_- \nabla_s \Psi^a- i\, \pa_s \bar\Psi_4\g^s \Psi^4+ i\, \bar\Psi_4 \g^s \pa_s \Psi^4\right]\nonumber \\
  &-\e_{abc}\Big[\pa_s z^b{\Psi^T}^a \g^t\Psi^c+\pa_t z^b{\Psi^T}^a(1+2\Pi_+)\Psi^c\Big]-\e^{abc}\Big[\pa_s \bar z_b{\bar\Psi}_a \g^t{\bar\Psi^T}_c-\pa_t\bar z_b {\bar\Psi}_a(1+2\Pi_-) {\bar\Psi^T}_c\Big]\nonumber\\ 
 &+4 {\Psi^T}^a \Pi_+ \slashed{\pa}\bar z_a \g^t \Psi^4-4 {\bar\Psi}_a \Pi_- \slashed{\pa}z^a \g^t {\bar\Psi^T}_4+ 4\, i\, (2\bar\Psi_a \g^t \Pi_+ \Psi^a+\bar \Psi_4 \g^t \Psi^4)\nabla_s x
\end{align}
where we introduced the short-hand notation $\nabla_s=\pa_s-1$ and the derivative is understood to act only on the nearest neighbor field. 

The quartic Lagrangian is slightly more involved
\begin{eqnarray}
 L_4=&8\,\phi^2 &\Big[4(\nabla_s x)^2+\pa_i\phi\pa^i \phi+\frac16 \phi^2+ \pa_i z^a \pa^i \bar z_a -2\, i\, \nabla_s \bar\Psi_a \Pi_+ \Psi^a-2\, i\, \bar\Psi_a \Pi_+ \nabla_s \Psi^a\nonumber\\
 %&\phantom{=8\,\phi^2\Big[}
 &&+ i\, \pa_s \bar\Psi_4 \g^s \Psi^4- i\, \bar\Psi_4 \g^s\pa_s \Psi^4\Big]\nonumber \\
 %&
 +&\frac13 &\Big[\bar z_a \pa_i z^a \bar z_c \pa^i z^c+ z^a \pa_i \bar z_a  z^c \pa^i \bar z_c-z^a \pa_i \bar z_a \bar z_c \pa_i z^c-|z|^2 \pa_i z^a \pa^i \bar z_a\Big] \nonumber\\
 %&
 +&4\,\phi &\Big[\e_{abc}\pa_s z^b{\Psi^T}^a\g^t\Psi^c+\e^{abc} \pa_s \bar z_b{\bar\Psi}_a\g^t{\bar\Psi^T}_c %\nonumber\\
 %&\phantom{-4\,\phi\Big[}
 %&
 -4 {\Psi^T}^a \Pi_+ \g^s \pa_s \bar z_a \g^t \Psi^4+4 {\bar\Psi}_a \Pi_- \g^s \pa_s z^a \g^t {\bar\Psi^T}_4\nonumber \\
 &&- 6\, i\, (2\bar\Psi_a \g^t \Pi_+ \Psi^a+\bar \Psi_4 \g^t \Psi^4)\nabla_s x\Big] \nonumber\\
 +&i &\Big[\bar \Psi_a (\bar z_b \slashed{\pa} z^a-z^a \slashed{\pa}\bar z_b)\Psi^b-\bar \Psi_a (\bar z_b \slashed{\pa} z^b-z^b \slashed{\pa}\bar z_b)\Psi^a+2\, \bar \Psi_4 (\bar z_a \slashed{\pa} z^a-z^a \slashed{\pa}\bar z_a)\Psi^4\Big] \nonumber \\
 %&
 +&8\, &\Big[(\bar\Psi_a\g^t\Pi_+ \Psi^a+\bar \Psi_4 \g^t \Pi_+ \Psi^4)^2+\e_{abc}{\Psi^T}^a \Pi_+ \Psi^b\ {\Psi^T}^c\Pi_+ \Psi^4+\e^{abc}{\bar\Psi}_a \Pi_- {\bar\Psi^T}_b\ {\bar\Psi}_c\Pi_- {\bar\Psi^T}_4 \nonumber \\
 &&-(\bar\Psi_4\g^t\Psi^4)^2\Big]
\end{eqnarray}
\vfill
\newpage

\bibliographystyle{JHEP}

\bibliography{biblio}

\end{document}